\input harvmac  
\input epsf  
 
%
%

\noblackbox 
%
 
 
\def\unlockat{\catcode`\@=11} 
\def\lockat{\catcode`\@=12} 
 
\unlockat 
 
\def\newsec#1{\global\advance\secno by1\message{(\the\secno. #1)} 
\global\subsecno=0\global\subsubsecno=0\eqnres@t\noindent 
{\bf\the\secno. #1} 
\writetoca{{\secsym} {#1}}\par\nobreak\medskip\nobreak} 
\global\newcount\subsecno \global\subsecno=0 
\def\subsec#1{\global\advance\subsecno 
by1\message{(\secsym\the\subsecno. #1)} 
\ifnum\lastpenalty>9000\else\bigbreak\fi\global\subsubsecno=0 
\noindent{\it\secsym\the\subsecno. #1} 
\writetoca{\string\quad {\secsym\the\subsecno.} {#1}} 
\par\nobreak\medskip\nobreak} 
\global\newcount\subsubsecno \global\subsubsecno=0 
\def\subsubsec#1{\global\advance\subsubsecno by1 
\message{(\secsym\the\subsecno.\the\subsubsecno. #1)} 
\ifnum\lastpenalty>9000\else\bigbreak\fi 
\noindent\quad{\secsym\the\subsecno.\the\subsubsecno.}{#1} 
\writetoca{\string\qquad{\secsym\the\subsecno.\the\subsubsecno.}{#1}} 
\par\nobreak\medskip\nobreak} 
  
\def\subsubseclab#1{\DefWarn#1\xdef 
#1{\noexpand\hyperref{}{subsubsection}%
{\secsym\the\subsecno.\the\subsubsecno}%
{\secsym\the\subsecno.\the\subsubsecno}}%
\writedef{#1\leftbracket#1}\wrlabeL{#1=#1}}
\lockat

\def\ie{{\it i.e.}}  
\def\eg{{\it e.g.}}


\font\manual=manfnt \def\dbend{\lower3.5pt\hbox{\manual\char127}}

\def\IZ{\relax\ifmmode\mathchoice  
{\hbox{\cmss Z\kern-.4em Z}}{\hbox{\cmss Z\kern-.4em Z}}  
{\lower.9pt\hbox{\cmsss Z\kern-.4em Z}}  
{\lower1.2pt\hbox{\cmsss Z\kern-.4em Z}}\else{\cmss Z\kern-.4em  
Z}\fi}  
\def\half{{1\over 2}}  
  
\def\p{\partial}  
\def\pb{\bar{\partial}}


\def\IZ{\relax\ifmmode\mathchoice  
{\hbox{\cmss Z\kern-.4em Z}}{\hbox{\cmss Z\kern-.4em Z}}  
{\lower.9pt\hbox{\cmsss Z\kern-.4em Z}}  
{\lower1.2pt\hbox{\cmsss Z\kern-.4em Z}}\else{\cmss Z\kern-.4em  
Z}\fi}  
\def\IB{\relax{\rm I\kern-.18em B}}  
\def\IC{{\relax\hbox{$\inbar\kern-.3em{\rm C}$}}}  
\def\ID{\relax{\rm I\kern-.18em D}}  
\def\IE{\relax{\rm I\kern-.18em E}}  
\def\IF{\relax{\rm I\kern-.18em F}}  
\def\IG{\relax\hbox{$\inbar\kern-.3em{\rm G}$}}  
\def\IGa{\relax\hbox{${\rm I}\kern-.18em\Gamma$}}  
\def\IH{\relax{\rm I\kern-.18em H}}  
\def\II{\relax{\rm I\kern-.18em I}}  
\def\IK{\relax{\rm I\kern-.18em K}}  
\def\IP{\relax{\rm I\kern-.18em P}}  
\def\IQ{\relax\hbox{$\inbar\kern-.3em{\rm Q}$}}

\def\inbar{\,\vrule height1.5ex width.4pt depth0pt}

\def\mod{{\rm mod}}  
\def\p{\partial}  
\def\pb{{\bar \p}}

\font\cmss=cmss10 \font\cmsss=cmss10 at 7pt  
\def\IR{\relax{\rm I\kern-.18em R}}

\def\Tr{\rm Tr}

%
%
  
\def\makeblankbox#1#2{\hbox{\lower\dp0\vbox{\hidehrule{#1}{#2}%
   \kern -#1
   \hbox to \wd0{\hidevrule{#1}{#2}%
      \raise\ht0\vbox to #1{}
      \lower\dp0\vtop to #1{}
      \hfil\hidevrule{#2}{#1}}%
   \kern-#1\hidehrule{#2}{#1}}}%
}%
\def\hidehrule#1#2{\kern-#1\hrule height#1 depth#2 \kern-#2}%
\def\hidevrule#1#2{\kern-#1{\dimen0=#1\advance\dimen0 by #2\vrule  
    width\dimen0}\kern-#2}%
\def\openbox{\ht0=1.2mm \dp0=1.2mm \wd0=2.4mm  \raise 2.75pt  
\makeblankbox {.25pt} {.25pt}  }

\def\bun#1/#2{\leavevmode  
   \kern.1em \raise .5ex \hbox{\the\scriptfont0 #1}%
   \kern-.1em $/$%
   \kern-.15em \lower .25ex \hbox{\the\scriptfont0 #2}%
}

\def\opensquare{\ht0=3.4mm \dp0=3.4mm \wd0=6.8mm  \raise 2.7pt  
\makeblankbox {.25pt} {.25pt}  }  
  
  
\def\sector#1#2{\ {\scriptstyle #1}\hskip 1mm  
\mathop{\opensquare}\limits_{\lower 1mm\hbox{$\scriptstyle#2$}}\hskip 1mm}  
  
\def\tsector#1#2{\ {\scriptstyle #1}\hskip 1mm  
\mathop{\opensquare}\limits_{\lower 1mm\hbox{$\scriptstyle#2$}}^\sim\hskip 1mm}  
  

\def\inbar{\,\vrule height1.5ex width.4pt depth0pt}  
  
\def\p{\partial}  
  
\def\pb{{\bar \p}}  
  
\font\cmss=cmss10 \font\cmsss=cmss10 at 7pt  
\def\IR{\relax{\rm I\kern-.18em R}}

\def\Tr{\rm Tr}

  
\def\sst{\scriptscriptstyle}  
  
\def\frac#1#2{{#1\over#2}}  
\def\coeff#1#2{{\textstyle{#1\over #2}}}  
\def\half{\frac12}  
\def\hf{{\textstyle\half}}

\def\inbar{\,\vrule height1.5ex width.4pt depth0pt}  
\def\IC{\relax\hbox{$\inbar\kern-.3em{\rm C}$}}  
\def\IR{\relax{\rm I\kern-.18em R}}  
\def\IP{\relax{\rm I\kern-.18em P}}

%
%
\catcode`\@=11  
\def\slash#1{\mathord{\mathpalette\c@ncel{#1}}}  
\overfullrule=0pt

\def\II{{\cal I}}

\def\LL{{\cal L}}

\def\RR{{\cal R}}

\def\ZZ{{\cal Z}}

\def\underrel#1\over#2{\mathrel{\mathop{\kern\z@#1}\limits_{#2}}}

\catcode`\@=12  
  
  
%

\def\det{{\rm det}}  
  
\def\mod{{\rm mod}}

\def\det{{\rm det}}  
\def\exp{{\rm exp}}

\def\st{\scriptstyle}  
\def\gop{g_{\rm op}}  
\def\gcl{g_{\rm cl}}  
\def\vt#1#2#3{ {\vartheta[{#1 \atop  #2}](#3\vert \tau)} }  
\def\ceff{c_{\rm eff}}  
\def\l{\ell}  
  
\lref\DixonQV{  
L.~J.~Dixon, D.~Friedan, E.~J.~Martinec and S.~H.~Shenker,  
``The Conformal Field Theory Of Orbifolds,''  
Nucl.\ Phys.\ B {\bf 282}, 13 (1987).  
}  
  
\lref\ZamolodchikovGT{  
A.~B.~Zamolodchikov,  
``'Irreversibility' Of The Flux Of The Renormalization Group In   
A 2-D Field Theory,''  
JETP Lett.\  {\bf 43}, 730 (1986)  
[Pisma Zh.\ Eksp.\ Teor.\ Fiz.\  {\bf 43}, 565 (1986)].  
}  
  
\lref\MartinecCF{  
E.~J.~Martinec and W.~McElgin,  
``String theory on AdS orbifolds,''  
hep-th/0106171.  
}  
  
\lref\LercheUY{  
W.~Lerche, C.~Vafa and N.~P.~Warner,  
``Chiral Rings In N=2 Superconformal Theories,''  
Nucl.\ Phys.\ B {\bf 324}, 427 (1989).  
}  
  
\lref\KutasovPF{  
D.~Kutasov,  
``Irreversibility of the renormalization   
group flow in two-dimensional quantum gravity,''  
Mod.\ Phys.\ Lett.\ A {\bf 7}, 2943 (1992)  
hep-th/9207064.  
}  
  
\lref\KutasovSV{  
D.~Kutasov and N.~Seiberg,  
``Number of degrees of freedom,   
density of states and tachyons in string theory and CFT,''  
Nucl.\ Phys.\ B {\bf 358}, 600 (1991).  
}  
  
\lref\AspinwallEV{  
P.~S.~Aspinwall,  
``Resolution of orbifold singularities in string theory,''  
hep-th/9403123.  
}  
  
\lref\CallanDJ{  
C.~G.~Callan, J.~A.~Harvey and A.~Strominger,  
``World sheet approach to heterotic instantons and solitons,''  
Nucl.\ Phys.\ B {\bf 359}, 611 (1991).  
}  
  
\lref\AdamsSV{  
A.~Adams, J.~Polchinski and E.~Silverstein,  
``Don't panic! Closed string tachyons in ALE space-times,''  
hep-th/0108075.  
}  
 
\lref\HarveyGQ{See for example the discussion in  
J.~A.~Harvey, S.~Kachru, G.~W.~Moore and E.~Silverstein, 
``Tension is dimension,'' 
JHEP {\bf 0003}, 001 (2000) 
hep-th/9909072. 
}  
 
\lref\HarveyNA{  
J.~A.~Harvey, D.~Kutasov and E.~J.~Martinec,  
``On the relevance of tachyons,''  
hep-th/0003101.  
}  
\lref\KutasovQP{  
D.~Kutasov, M.~Marino and G.~W.~Moore,  
``Some exact results on tachyon condensation in string field theory,''  
JHEP {\bf 0010}, 045 (2000)  
hep-th/0009148.  
}  
\lref\KutasovAQ{  
D.~Kutasov, M.~Marino and G.~W.~Moore,  
``Remarks on tachyon condensation in superstring field theory,''  
hep-th/0010108.  
}  
  
\lref\AffleckTK{  
I.~Affleck and A.~W.~Ludwig,  
``Universal noninteger 'ground state degeneracy' in critical   
quantum systems,'' Phys.\ Rev.\ Lett.\  {\bf 67}, 161 (1991).  
}  
\lref\MarinoQC{  
M.~Marino,  
``On the BV formulation of boundary superstring field theory,''  
JHEP {\bf 0106}, 059 (2001)  
hep-th/0103089.  
}  
\lref\NiarchosSI{  
V.~Niarchos and N.~Prezas,  
``Boundary superstring field theory,''  
hep-th/0103102.  
}  
  
\lref\CecottiTH{  
S.~Cecotti and C.~Vafa,  
``Massive orbifolds,''  
Mod.\ Phys.\ Lett.\ A {\bf 7}, 1715 (1992)  
hep-th/9203066.  
}  
  
\lref\ZamolodchikovDB{  
A.~B.~Zamolodchikov,  
``Conformal Symmetry And Multicritical Points In Two-Dimensional   
Quantum Field Theory,''  
Sov.\ J.\ Nucl.\ Phys.\  {\bf 44}, 529 (1986)  
[Yad.\ Fiz.\  {\bf 44}, 821 (1986)].  
}  
\lref\KastorEF{  
D.~A.~Kastor, E.~J.~Martinec and S.~H.~Shenker,  
``RG Flow In N=1 Discrete Series,''  
Nucl.\ Phys.\ B {\bf 316}, 590 (1989).  
}  
  
\lref\ColemanTJ{  
S.~R.~Coleman,  
``Why There Is Nothing Rather Than Something:   
A Theory Of The Cosmological Constant,''  
Nucl.\ Phys.\ B {\bf 310}, 643 (1988).  
}  
  
\lref\szenes{A. Szenes, ``The combinatorics of the Verlinde   
formulas,''  alg-geom/9402003}  
  
\lref\GinspargIS{  
P.~Ginsparg and G.~W.~Moore,  
``Lectures On 2-D Gravity And 2-D String Theory,''  
hep-th/9304011.  
}  
  
\lref\DiFrancescoNW{  
P.~Di Francesco, P.~Ginsparg and J.~Zinn-Justin,  
``2-D Gravity and random matrices,''  
Phys.\ Rept.\  {\bf 254}, 1 (1995)  
hep-th/9306153.  
}  
  
\lref\OoguriWJ{  
H.~Ooguri and C.~Vafa,  
``Two-Dimensional Black Hole and Singularities of CY Manifolds,''  
Nucl.\ Phys.\ B {\bf 463}, 55 (1996)  
hep-th/9511164.  
}  
  
\lref\DouglasSW{  
M.~R.~Douglas and G.~W.~Moore,  
``D-branes, Quivers, and ALE Instantons,''  
hep-th/9603167.  
}  
  
\lref\GiveonPX{  
A.~Giveon and D.~Kutasov,  
``Little string theory in a double scaling limit,''  
JHEP {\bf 9910}, 034 (1999)  
hep-th/9909110.  
}  
  
\lref\GiveonTQ{  
A.~Giveon and D.~Kutasov,  
``Comments on double scaled little string theory,''  
JHEP {\bf 0001}, 023 (2000)  
hep-th/9911039.  
}  
  
\lref\GiveonZM{  
A.~Giveon, D.~Kutasov and O.~Pelc,  
``Holography for non-critical superstrings,''  
JHEP {\bf 9910}, 035 (1999)  
hep-th/9907178.  
}  
  
\lref\KutasovJP{  
D.~Kutasov and D.~A.~Sahakyan,  
``Comments on the thermodynamics of little string theory,''  
JHEP {\bf 0102}, 021 (2001)  
hep-th/0012258.  
}  
  
\lref\AharonyUB{  
O.~Aharony, M.~Berkooz, D.~Kutasov and N.~Seiberg,  
``Linear dilatons, NS5-branes and holography,''  
JHEP {\bf 9810}, 004 (1998)  
hep-th/9808149.  
}  
 
\lref\TseytlinQB{ 
A.~A.~Tseytlin, 
``Magnetic backgrounds and tachyonic instabilities  
in closed string  theory,'' 
hep-th/0108140. 
}

\lref\DabholkarGZ{  
A.~Dabholkar,  
``On condensation of closed-string tachyons,''  
hep-th/0109019.  
}  
 
\lref\SuyamaGD{ 
T.~Suyama, 
``Properties of string theory on Kaluza-Klein Melvin background,'' 
hep-th/0110077. 
} 
 
\lref\RussoNA{ 
J.~G.~Russo and A.~A.~Tseytlin, 
``Supersymmetric fluxbrane intersections and closed string tachyons,'' 
hep-th/0110107. 
} 
 
\lref\TakayanagiJJ{
T.~Takayanagi and T.~Uesugi,
``Orbifolds as Melvin geometry,''
hep-th/0110099.
}

\lref\VafaRA{ 
C.~Vafa, 
``Mirror Symmetry and Closed String Tachyon Condensation,'' 
hep-th/0111051. 
}

\lref\GerasimovZP{  
A.~A.~Gerasimov and S.~L.~Shatashvili,  
``On exact tachyon potential in open string field theory,''  
JHEP {\bf 0010}, 034 (2000)  
hep-th/0009103.  
}  
  
\lref\GiddingsCG{  
S.~B.~Giddings and A.~Strominger,  
``Axion Induced Topology Change In Quantum Gravity And String Theory,''  
Nucl.\ Phys.\ B {\bf 306}, 890 (1988).  
}  
  
\lref\SenMD{  
A.~Sen,  
``Supersymmetric world-volume action for non-BPS D-branes,''  
JHEP {\bf 9910}, 008 (1999)  
hep-th/9909062.  
}  
  
\lref\YiHD{  
P.~Yi,  
``Membranes from five-branes and fundamental strings from Dp branes,''  
Nucl.\ Phys.\ B {\bf 550}, 214 (1999)  
hep-th/9901159.  
}  
  
\lref\BergmanXF{  
O.~Bergman, K.~Hori and P.~Yi,  
``Confinement on the brane,''  
Nucl.\ Phys.\ B {\bf 580}, 289 (2000)  
hep-th/0002223.  
}  
  
\lref\KlebanPF{  
M.~Kleban, A.~E.~Lawrence and S.~H.~Shenker,  
``Closed strings from nothing,''  
Phys.\ Rev.\ D {\bf 64}, 066002 (2001)  
hep-th/0012081.  
}  
 
\lref\bpv{W. Barth, C. Peters, and A. van de Ven, 
{\it Compact Complex Surfaces},  
Ergebnisse der Mathematik und ihrer Grenzgebiete (3)  
[Results in Mathematics and Related Areas (3)], 4.  
Springer-Verlag, Berlin, 1984.} 
 
\lref\fulton{W. Fulton,  
{\it Introduction to toric varieties}, 
Annals of Mathematics Studies, 131.  
Princeton University Press, Princeton, NJ, 1993.} 
  
\lref\SenMG{ 
A.~Sen, 
``Non-BPS states and branes in string theory,'' 
hep-th/9904207. 
} 
 
\lref\hkmmtwo{J. Harvey, D. Kutasov, E. Martinec, and G. Moore, 
work in progress.} 
 
\lref\oda{T. Oda, {\it Convex Bodies and Algebraic Geometry}, Springer-Verlag, 
Berlin, 1988.} 
 
\lref\agm{ 
P.~S.~Aspinwall, B.~R.~Greene and D.~R.~Morrison, 
``Calabi-Yau moduli space, mirror manifolds and  
spacetime topology  change in string theory,'' 
Nucl.\ Phys.\ B {\bf 416}, 414 (1994) 
hep-th/9309097. 
} 
 
\lref\hkt{ 
S.~Hosono, A.~Klemm and S.~Theisen, 
``Lectures on mirror symmetry,'' 
hep-th/9403096. 
} 
 
\lref\MorrisonPlesser{ 
D.~R.~Morrison and M.~Ronen Plesser, 
``Summing the instantons: Quantum cohomology and mirror symmetry in toric varieties,'' 
Nucl.\ Phys.\ B {\bf 440}, 279 (1995) 
hep-th/9412236. 
} 
 
\lref\GreeneCY{ 
B.~R.~Greene, 
``String theory on Calabi-Yau manifolds,'' 
hep-th/9702155. 
} 
 
\lref\Skarke{ 
H.~Skarke, 
``String dualities and toric geometry: An introduction,'' 
hep-th/9806059. 
} 
 
\lref\DiaconescuWY{ 
D.~E.~Diaconescu, G.~W.~Moore and E.~Witten, 
``E(8) gauge theory, and a derivation of K-theory from M-theory,'' 
hep-th/0005090. 
} 
 
\lref\MooreGB{ 
G.~W.~Moore and E.~Witten, 
``Self-duality, Ramond-Ramond fields, and K-theory,'' 
JHEP {\bf 0005}, 032 (2000) 
hep-th/9912279. 
} 
 
\lref\danilov{V.I. Danilov, ``The geometry of toric varieties,'' Russian 
Math. Surveys, {\bf 33} (1978) 97}  

\lref\CostaIF{
M.~S.~Costa, C.~A.~Herdeiro and L.~Cornalba,
``Flux-branes and the dielectric effect in string theory,''
hep-th/0105023.
}

\lref\SaffinKY{
P.~M.~Saffin,
``Gravitating fluxbranes,''
Phys.\ Rev.\ D {\bf 64}, 024014 (2001)
gr-qc/0104014.
}

\lref\DabholkarAI{
A.~Dabholkar,
``Strings on a cone and black hole entropy,''
Nucl.\ Phys.\ B {\bf 439}, 650 (1995)
hep-th/9408098.
}


\rightline{EFI-01-50}  
\rightline{RUNHETC-2001-35}  
\Title{  
\rightline{hep-th/0111154}}  
{\vbox{\centerline{Localized Tachyons and RG Flows}}}  
\bigskip  
\centerline{Jeffrey A. Harvey$^1$, David Kutasov$^1$,  
Emil J. Martinec$^1$ and Gregory Moore$^2$}  
\bigskip  
\centerline{$^1$ {\it Enrico Fermi Inst. and Dept. of Physics,  
University of Chicago}}  
\centerline{\it 5640 S. Ellis Ave., Chicago, IL 60637-1433, USA}  
\bigskip  
\centerline{$^2$ {\it Department of Physics, Rutgers University}}  
\centerline{\it Piscataway, NJ 08855-0849, USA}

\bigskip  
\noindent  
We study condensation of closed string tachyons living  
on defects, such as orbifold fixed planes and Neveu-Schwarz  
fivebranes. We argue that the high energy density of   
localized states decreases in the process of condensation of   
such tachyons. In some cases this means that $c_{\rm eff}$   
decreases along the flow; in others, $c_{\rm eff}$ remains   
constant and the decreasing quantity is a closed string analog,  
$g_{\rm cl}$, of the ``boundary entropy'' of D-branes.   
We discuss the non-supersymmetric orbifolds $\IC/\IZ_n$ and  
$\IC^2/\IZ_n$. In the first case tachyon condensation decreases 
$n$  and in some cases connects type II and type 0 vacua. In the 
second case non-singular orbifolds are related by tachyon 
condensation to both singular and non-singular ones. We verify 
that $g_{\rm cl}$ decreases in flows between non-singular orbifolds. 
The main tools in the analysis are the structure of the chiral 
ring of the perturbed theory, the geometry of the resolved 
orbifold singularities, and the throat description of singular 
conformal field theories.  
  
\vfill  
  
\Date{November 16, 2001}


\newsec{Introduction}  
  
In recent years significant progress    
has been achieved in string theory from an analysis of  
string propagation in the presence of defects  
such as D-branes, NS5-branes and orbifolds.   
The study of impurities in string theory led to important insights   
into fundamental issues (\eg\ the AdS/CFT correspondence)   
and might be relevant to phenomenology   
(for example in the framework of ``brane worlds'').  
In such backgrounds there are two kinds of excitations,  
those that are localized on the defects,   
and those that propagate everywhere in spacetime.  
In the D-brane case, the localized states are open strings ending   
on the branes, while the delocalized ones correspond to closed   
strings. For orbifolds, they are twisted and untwisted  
sector excitations, respectively.  
  
In the absence of spacetime supersymmetry, string vacua   
typically contain tachyons, and it   
is of interest to analyze their condensation. This requires  
an off-shell formulation of the theory, and is in general  
not well understood. It is believed that in general tachyon   
condensation drastically modifies the geometry of spacetime,   
and even its dimensionality.

For vacua with impurities, one can consider systems   
in which supersymmetry is only broken by the impurity.   
In such situations, one generically expects to find tachyons   
living on the impurity (``localized tachyons''), and their   
condensation might be under better control than the general  
case. For D-branes, this was widely discussed in the past two  
years following the work of Sen \SenMG.  Here we will focus on   
closed string defects, such as orbifolds and NS5-branes.   
This case was recently studied in \AdamsSV\ (see also \MartinecCF),  
where the similarity of this problem to open string tachyon  
condensation on unstable D-brane systems was noted. We will continue this study  
using techniques that can be thought of as a direct generalization   
of \refs{\HarveyNA,\GerasimovZP,\KutasovQP,\KutasovAQ} to closed strings.  
  
We will see that, as in the open string case, the worldsheet   
renormalization group (RG)  
provides a simple conceptual framework for understanding localized  
tachyon condensation.  It also leads to some quantitative constraints   
on possible decays.  A nice feature of the analysis is a uniform   
treatment of all localized defects in weakly coupled string theory.  
  
A characteristic feature of the renormalization group  
in two dimensions is a decrease along flows of   
the asymptotic high energy density of states  
\eqn\asdens{  
  \rho(E)\sim \half g\;(\ceff/3E^3)^{1\over 4}\; 
	\exp\left[2\pi\sqrt{\ceff E/3}\right]\ .}  
A number of examples of this phenomenon are known:  
\item{1.} Unitary compact\foot{By a compact CFT, we mean  
one with a discrete spectrum of scaling dimensions.}  
conformal field theories.  
Here, $\ceff=c$, the Virasoro central charge; the decrease  
of $c$ along RG flows in this case was proven by  
Zamolodchikov \ZamolodchikovGT.   
\item{2.} All compact CFT's, perhaps non-unitary,  
in which all states contribute positively to the torus partition sum.  
Then one has \KutasovSV  
\eqn\ceffdef{  
  \ceff=-\frac{6E_{\rm min}}{\pi}=c-24 h_{\rm min}\ ,  
}  
where $h_{\rm min}$ is the lowest scaling dimension in the theory  
and $c_{\rm eff}$ is believed to decrease along RG flows (this has  
not been proven in general; see \eg\ \refs{\KutasovSV,\KutasovPF} for 
discussions).  
Examples where RG flows of such systems have been studied  
include the non-unitary minimal models, for which  
\eqn\cmin{  
  c=1-\frac{6(p-p')^2}{pp'}\quad,\qquad \ceff=1-\frac{6}{pp'}\quad;  
}  
RG flows decrease $p$ and/or $p'$   
\refs{\DiFrancescoNW,\GinspargIS}, and hence $\ceff$.  
\item{3.} On worldsheets with boundary,  
RG flows by boundary perturbations cannot affect the  
central charge; in this case, it is the open string analog 
of the coefficient $g$ in \asdens\ (which is related to 
the ``boundary entropy'' of \AffleckTK)  
which decreases along flows \refs{\AffleckTK,\KutasovQP}.  
  
\noindent  
One of the main purposes of this paper is to present another  
class of examples of the decrease of the density of states  
\asdens\ along RG flows. Following \AdamsSV, we will study  
non-compact sigma-models with localized states associated  
with impurities that break (spacetime) supersymmetry. An   
important class of examples are the non-compact orbifolds   
$\IR^m/\Gamma$. Typically, such models contain localized  
(twisted sector) tachyons and one can ask what happens  
to the density of states \asdens\ as these condense. It is  
natural in this case to define $\rho(E)$ as the density of  
{\it localized} states, which are normalizable, as opposed  
to the delocalized states which are only delta-function  
normalizable (\ie\ non-normalizable in infinite volume).   
  
It was proposed in \AdamsSV\ that non-compact orbifold models  
often flow in the IR to other orbifold models. As we review  
in section 2, $c_{\rm eff}$ does not change in such flows.  
Therefore, we propose that in these cases the quantity that  
decreases along the RG flow is   
the coefficient $g$ in \asdens, which we denote by $\gcl$  
to distinguish it from its open string counterpart $\gop$.  
  
Since we have not proven the ``$g_{\rm cl}$-conjecture,''  
in the rest of the paper we present a number of examples  
of RG flows in orbifold theories, and test the conjecture  
for them. In order to analyze  
the far infrared physics reached by twisted sector  
tachyon perturbations, we restrict to  
flows that preserve $N=2$ worldsheet supersymmetry.  
In these cases, the chiral ring \LercheUY\  
provides a powerful diagnostic.  The restriction  
to $N=2$ supersymmetry is not a severe limitation;  
as we will see, it allows for many interesting flows.  
The picture obtained  
is compatible with the ``$\gcl$-conjecture''.  
  
NS5-branes are another interesting 
class of defects that are  related to orbifolds  
but exhibit somewhat different physics.  
String perturbation theory typically breaks down  
in the presence of coincident fivebranes due to the appearance  
of an infinite throat \CallanDJ, but there are examples in which  
the system is weakly coupled (obtained by going to the Coulomb branch  
of the theory, where the throat is capped \GiveonPX). We show that  
in these systems $c_{\rm eff}$ typically decreases along RG flows  
(and thus $g$ in \asdens\ is superfluous).   
  
The plan of the paper is the following. In section 2  
we discuss non-compact orbifolds. We show that the infinite  
volume limit serves as a sort of decoupling limit between  
twisted and untwisted states. In particular, we show that   
the central charge does not change under twisted sector  
perturbations. We introduce $g_{\rm cl}$ and give a general  
formula for it. We also show that $g_{\rm cl}$ does not change 
along moduli spaces corresponding to twisted sector marginal 
operators. 
  
In section 3 we discuss the special case of $\IC/\IZ_n$. We analyze  
the condensation of twisted tachyons that preserves $N=2$ worldsheet  
supersymmetry, both for type 0 and type II strings, by studying  
the perturbed chiral rings. We find that tachyon condensation  
relates string theory on $\IR^{7,1}\times\IC/\IZ_n$ to a direct sum   
of disconnected cones of the form $\IR^{7,1}\times\IC/\IZ_{n_i}$ with  
$n_i<n$. For the type II theory, one of these cones has type II  
strings living on it, while the rest are type 0 vacua. We also compute  
$g_{\rm cl}$ and show that this process is compatible with the   
$g_{\rm cl}$-conjecture.  
  
In section 4 we discuss the $\IC^2/\IZ_n$ case. Subsection 4.1 discusses 
the Hirzebruch-Jung theory of singularity resolution and its relation to 
the chiral ring in the orbifold SCFT. In subsection 4.2 
we review the relation between  
spacetime supersymmetric orbifolds and a vacuum with  
NS5-branes on $\IR^3\times S^1$. In the rest of section 4 we use  
these tools to analyze the RG flows in various examples.  
  
Section 5 contains some applications of the ideas developed here to RG 
flows in fivebrane theories with a non-compact transverse space. We 
end in section 6 with a discussion of our results and further  
applications. Three appendices contain technical results on the  
computation of $g_{\rm cl}$, gauge couplings on D-brane probes, and  
the structure of chiral rings for certain orbifolds of $\IC^2$.  
  
Condensation of localized closed string tachyons has also been studied   
recently in 
\refs{\TseytlinQB,\DabholkarGZ,\SuyamaGD,\RussoNA,\TakayanagiJJ,\VafaRA}.  
  

\newsec{Localized closed string tachyon condensation and $g_{\rm cl}$}  
  
A prototype of the problem we will study is the following. Consider  
string propagation in the spacetime  
\eqn\aaa{\IR^{d-1,1}\times\IR^{10-d}/\Gamma\ ,}  
where $\Gamma$ is a discrete subgroup of $SO(10-d)$, whose action   
on $\IR^{10-d}$ has a single fixed point (say $\vec y=0$, with   
$\vec y\in \IR^{10-d}$). Twisted sector states give rise to fields  
localized at $\vec y=0$ (which thus live in $d$ spacetime dimensions),  
while untwisted states propagate in the full ten dimensional spacetime.  
Tachyons typically appear when $\Gamma$ is not in $SU(5-[(d+1)/2])$ 
(or $G_2$ or $Spin(7)$ for $d=3,2$),  
so supersymmetry is broken. Here  $[x]$ denotes the integer part of $x$. 
  
``Localized instabilities'' in \aaa\ correspond to twisted sector  
scalars which are tachyonic (or massless, with a higher order unstable  
potential). Such scalars will condense, and it is of interest to   
determine the endpoint of this process.   
  
In the open string case, it is useful to think about this problem  
in terms of the worldsheet RG \HarveyNA. In that case, the   
instabilities are associated with open string tachyons. Condensing   
them corresponds to studying the theory in the presence of the   
relevant boundary vertex operators in the worldsheet action. The   
endpoint of condensation corresponds to the IR fixed point of this flow.   
  
Similarly here, condensation of localized closed string tachyons   
(which correspond to relevant perturbations in the twisted sector   
of the worldsheet CFT) can be studied by following the RG flow of   
the worldsheet theory perturbed by the tachyon vertex operators. By   
analogy  with the D-brane example, one expects to be able to define   
an off-shell quantity which is monotonically decreasing along the   
flow and at the fixed points of the RG measures the number of  
localized degrees of freedom.  
  
In the open string case, this quantity is the ``boundary entropy''   
$g$, whose decrease along RG flows was conjectured by Affleck and   
Ludwig \AffleckTK\ and proven in \KutasovQP\ (see also \GerasimovZP,  
and \refs{\KutasovAQ,\MarinoQC,\NiarchosSI} for the worldsheet  
supersymmetric case), where it was also shown that this quantity  
is the off-shell action for open string excitations.   
One way of defining $g$, which makes manifest its   
interpretation as the number of boundary degrees of freedom,   
is via the density of open string states\foot{In situations  
where there are different sectors of open strings, we have in mind 
the total density of states, summed over the different sectors.} 
at high energy, $\rho(E\to\infty)$:  
\eqn\bbb{\rho_{\rm open}(E)\sim  
\half g_{\rm op}\;(\ceff/6E^3)^{{1\over4}} \; 
\exp\left[2\pi\sqrt{\ceff E/6}\right]\ ,}  
where $E$ is the eigenvalue of $L_0-(c/24)$. 
Since the central charge does not change under boundary RG flow,   
$g_{\rm op}$ is the leading measure of the change in the number   
of open string degrees of freedom of the system in such flows.  
One can show that $g_{\rm op}$ is related to the boundary 
entropy $g$ via $g_{\rm op}=g^2$ \AffleckTK.   
  
For non-supersymmetric orbifolds of the form \aaa, one might    
expect that the quantity that decreases along the RG when the   
theory is perturbed by twisted sector relevant operators is   
the central charge $c$. Indeed, Zamolodchikov's $c$-theorem   
\ZamolodchikovGT\ asserts that in unitary theories, the central   
charge decreases along RG flows. We next briefly review the   
argument and discuss the subtleties with it in this case.  
  
Zamolodchikov has shown that  
\eqn\ccc{{d c\over dt}=-12|z|^4\langle\Theta(z)\Theta(0)\rangle\ ,}  
where $t=2\ln|z|$ is the log of the RG scale, $c$ is the   
scale-dependent central charge  
\eqn\ddd{c(t)=2z^4\langle T(z)T(0)\rangle-4z^3\bar z  
\langle T(z)\Theta(0)\rangle-6|z|^4\langle \Theta(z)\Theta(0)\rangle\ ,}  
and $\Theta=T_{z\bar z}$ is the trace of the stress tensor. At fixed  
points of the RG, $\Theta$ vanishes and $c$ reduces to the standard  
Virasoro central charge. More generally,  one has  
\eqn\eee{\Theta=\beta^i\phi_i}  
where $\beta^i$ are the $\beta$-functions for the couplings   
$\lambda^i$, and $\phi_i$ are the corresponding perturbations.   
Plugging \eee\ in \ccc\ one has  
\eqn\fff{{d c\over dt}=-12\beta^i\beta^jG_{ij}\ ,}  
where  
\eqn\ggg{G_{ij}=\langle\phi_i(z)\phi_j(0)\rangle|z|^4}  
is the Zamolodchikov metric. In a unitary QFT, the  
metric $G$ is expected to be positive definite, and  
thus $c$ decreases along RG flows.  
  
In \AdamsSV\ it was argued that in the non-compact orbifold   
theories \aaa\ this reasoning might fail, and the central   
charge does not in fact change along RG flows associated   
with twisted sector perturbations. We next discuss how   
these statements can be reconciled with the $c$-theorem.   
  
It is important that, as mentioned above, in the orbifolds   
\aaa\ there are two kinds of excitations. Vertex operators   
that belong to the untwisted sector of the orbifold describe   
excitations that propagate in the full ten dimensional spacetime,   
while twisted sector states are restricted to the $d$ dimensional  
manifold $\vec y=0$. This has important implications for the   
correlation functions of the theory.    
  
Denote by $U$ general untwisted sector Virasoro primaries and by   
$T$ twisted ones. Consider correlation functions of the form  
\eqn\tuntcor{  
\langle U_1(z_1)\cdots U_n(z_n)T_1(w_1)\cdots T_m(w_m)\rangle\ .}  
Such correlation functions are given by the path integral over   
the worldsheet fields on the target space \aaa. As is standard   
in QFT, the correlation functions are normalized by dividing by   
the partition sum. There is an important difference between the   
behavior of correlation functions of untwisted operators (\ie\   
those with $m=0$ in \tuntcor), and correlation functions which   
contain twisted operators ($m\not=0$). The difference has to do   
with the volume dependence of the correlators, and is thus  
easiest to explain by replacing $R^{10-d}/\Gamma$ in   
\aaa\ by a compact orbifold with the same singularity, \eg\   
$S^{10-d}/\Gamma$, where $S^{10-d}$ is a round sphere of volume  
$V_{10-d}$. The curvature of the sphere breaks conformal symmetry,  
but since one is only using $V_{10-d}$ as a regulator, this can  
be ignored. Also, there might be more than one fixed point of   
$\Gamma$ on the sphere; again, this can be ignored in the large  
volume limit.  
  
We would like to analyze the dependence of the correlation functions   
\tuntcor\ on $V_{10-d}$ in the large volume limit.  
Before dividing by the partition sum $\ZZ$, untwisted correlation   
functions are proportional to the volume -- the vertex operators   
$U_i$ describe particles that live on the entire $S^{10-d}$.   
Since $\ZZ$ is also proportional to $V_{10-d}$, the normalized  
correlators \tuntcor\ with $m=0$ are independent of $V_{10-d}$,  
\eqn\norvun{\langle U_1\cdots U_n\rangle\simeq V_{10-d}^0\ .}  
On the other hand, (connected) correlation functions involving twist   
fields correspond to processes localized at the fixed point $\vec y=0$,   
and therefore the normalized correlators go for $m>0$ like  
\eqn\norvtw{  
\langle U_1\cdots U_nT_1\cdots T_m\rangle\simeq {1\over V_{10-d}}\ .}  
These properties may be inferred from the explicit computation  
of twist operator correlation functions in \DixonQV.  
Note that although one can rescale the operators $U_i$, $T_a$ by   
powers of $V_{10-d}$, \norvun, \norvtw\ are meaningful, since one   
can fix the normalization by requiring that correlators   
like \norvun\ scale as {\it the same} power of $V_{10-d}$   
for all $n$ (and similarly for \norvtw). This is a reasonable   
requirement to impose if one is planning on taking the large   
$V_{10-d}$ limit.   
  
Equations \norvun, \norvtw\ imply that the OPE's are regular in the   
limit $V_{10-d}\to\infty$:  
\eqn\untwope{\eqalign{  
U_iU_j=&C_{ij}^kU_k\cr  
T_aT_b=&D_{ab}^cT_c+B_{ab}^kU_k\cr  
}}  
with the structure constants $C_{ij}^k$, $D_{ab}^c$ finite in the   
large volume limit and $B_{ab}^k\sim 1/V_{10-d}$. Interestingly,   
as $V_{10-d} \to\infty$, all (normalized) correlation functions of   
the properly normalized twist operators go to zero \norvtw, while   
their OPE's remain finite \untwope. It is also clear that in the  
presence of twisted sector perturbations, \fff\ does not imply that   
$c$ is decreasing. Indeed, the Zamolodchikov metric \ggg,   
$G_{ab}=|z|^4\langle T_a(z) T_b(0)\rangle$, vanishes to all orders in   
conformal perturbation theory in the twisted couplings and thus $c$  
is constant along the flows.   
  
We now return to the question of what quantity changes along the   
twisted RG flows discussed above. We would like to propose that   
it is a closed string analog of \bbb, the high energy density  
of localized states. This density%
\foot{In this paper, we study the density of states in
{\it conformal field theory} on \aaa.  In string theory
on this spacetime, one has to further impose the
physical state condition $L_0=\bar L_0$ (see \KutasovSV\
for a related discussion).}
grows for high energy  
$E=L_0+\bar L_0-(c/12)$ as  
\eqn\mmm{  
  \rho(E)\sim \half g_{\rm cl}\;(c/3E^3)^{1\over 4}\; 
	\exp\left[2\pi\sqrt{cE/3}\right]\ .}  
As we just saw, the leading term in $\rho(E)$, which is governed  
by the central charge, is unchanged along the flow. Therefore,  
the leading measure of the density of states that {\it can} change,  
is the prefactor $g_{\rm cl}$. Since the number of degrees of  
freedom is expected to decrease, and in analogy to the open string  
case, we conjecture that along RG flows of the orbifold CFT perturbed   
by twisted vertex operators one has  
\eqn\nnn{g_{\rm cl}({\rm UV})>g_{\rm cl}({\rm IR})\ .}  
Note that equation \mmm\ defines $g_{\rm cl}$ at fixed points of the RG   
(as was done in \AffleckTK\ for $g_{\rm op}$). It would be interesting   
to define $g_{\rm cl}$ throughout the RG flow, and prove \nnn. One   
expects to be able to define such a quantity away from the fixed   
points since there should exist a classical off-shell spacetime   
effective action for localized states. This relies on the fact   
that these states are described by a non-gravitational theory;   
it is believed that for gravitational theories one should not   
be able to go off-shell (due to holography).   
  
Just like in the open string case \AffleckTK, one can relate $g_{\rm cl}$  
to a ``non-integer ground state degeneracy'' by using modular  
invariance of the one loop partition sum. Indeed, consider  
the partition sum for localized states  
\eqn\ztwist{  
\ZZ_{\rm tw}(\tau,\bar\tau)=\sum_{g\ne 1}  
{\rm Tr}_g [q^{L_0-{c\over 24}} \bar q^{\bar L_0-{c\over 24}}]\ ,}  
where the sum over $g$ runs over the different twisted sectors  
of the orbifold (the trace is over $\Gamma$-invariant states),   
and we will take the modular parameter $q$  
to be real, $q=\exp(-2\pi\tau_2)$. The high energy density of  
states \mmm\ determines the behavior of the partition sum  
\ztwist\ in the limit $\tau_2\to 0$. Indeed, plugging the density  
of states $\rho(E)$ \mmm\ into \ztwist, replacing the sum in the trace  
by an integral, and performing it by saddle point, one finds  
\eqn\zgcl{ 
\ZZ_{\rm tw}(\tau_2\to 0)\sim g_{\rm cl}\;\exp(\pi c/6\tau_2 )\ .  
}  
The modular transformation $\tau_2\to1/\tau_2$ relates this  
to the contribution of the vacuum state to the torus partition  
sum at large $\tau_2$.   
  
Consider, for example, the orbifold with a diagonal modular  
invariant. The twisted partition sum in this case has the form   
\eqn\zfullorb{  
  \ZZ_{\rm tw}= {1\over \vert \Gamma\vert}\sum_{g\not=1}  
	\sum_{h\in\Gamma  \atop gh=hg}   
	\ZZ\Bigl(\sector{h}{g}\Bigr)\   
} 
where   
\eqn\zzoorr{\ZZ\Bigl(\sector{h}{g}\Bigr)={\rm Tr}_g h q^{L_0-{c\over24}} 
\bar q^{\bar L_0-{c\over24}},} 
the partition sum in the sector twisted by $g$ with an insertion 
of the operator representing $h$ on the Hilbert space. 
The sum over $h\in\Gamma$ in \zfullorb\ imposes the $\Gamma$  
invariance of the physical states. We are interested in the  
behavior of $\ZZ_{\rm tw}$ in the limit $\tau_2\to 0$.  
Under the modular transformation $\tau\to-1/\tau$, 
\eqn\modtransfz{ 
  \ZZ\Bigl(\sector{1}{g}\Bigr)(-{1\over\tau})= 
	\ZZ\Bigl(\sector{g}{1}\Bigr)(\tau)= 
	{1 \over \vert \det (1-\RR(g)) \vert^2}  
	\ZZ'\Bigl(\sector{g}{1}\Bigr)(\tau)~. 
} 
Here $\ZZ'$ is the partition sum with the zero modes  
excluded, and the inverse determinant factor comes  
from the zero mode integral, 
\eqn\zeromodeint{\int d^{10-d}p\;\delta^{10-d}(gp-p)= 
{1 \over \vert \det (1-\RR(g)) \vert^2}.} 
$\RR(g)$ is the rotation matrix representing $g$  
on the string coordinates. Taking the limit $\tau_2\to\infty$,  
one finds that 
\eqn\gengee{  
\gcl = {1\over \vert \Gamma\vert } \sum_{g\not=1}  
  {1 \over \vert \det (1-\RR(g)) \vert^2} . 
}  
We will evaluate this expression in particular examples below.  
Note that for compact orbifolds, the factor  
$\vert \det (1-\RR(g)) \vert^2$ gives the number of fixed points  
in the sector twisted by $g$, and the more familiar form of the  
modular transformation \modtransfz\ is obtained by multiplying  
\modtransfz\ by the number of fixed points. We are studying 
a non-compact orbifold with a single fixed point, which is the  
origin of the inverse determinant appearing in \modtransfz, \gengee.  
 
  
An intriguing fact is that \gengee\ is essentially the same  
as the $\eta$ invariant for the Dirac operator in the round metric  
on Lens spaces of the type $\IC^{10-d}/\Gamma$. One possible 
interpretation of this fact is that twisted sector strings can 
be thought of as random walks constrained to begin and end at 
the fixed point. For high energies, the random walk explores 
the region far from the fixed point, and most of the entropy 
of such strings comes from this region. This explains the  
role of $S^{10-d}/\Gamma$ in evaluating the asymptotic density 
of states. It also clarifies why $g_{\rm cl}$ is related to 
a boundary contribution to an index: it should be invariant 
under small deformation in the vicinity of the tip of the cone, 
since such deformations are not expected to influence the high 
energy density of states.  
 
The complexification of $\IR^{10-d}$  
to $\IC^{10-d}$ in this interpretation is perhaps due  
to the need to consider both left and right-moving degrees of  
freedom. It is also natural to expect that a relation between $g_{cl}$ 
and $\eta$ invariants can be found by combining the elliptic 
genus with the index theorem for manifolds with boundary. 
 
An example of a class of  
small perturbations under which $g_{\rm cl}$ should 
be invariant is deformations by exactly marginal twisted sector  
operators. These give rise to moduli spaces of CFT's which  
correspond to resolved orbifold singularities. Since the  
asymptotic geometry of the non-compact orbifold 
does not depend on the moduli, one would expect $g_{\rm cl}$ to 
remain constant along such moduli spaces (at least for finite 
distance in the space of CFT's). We next show that this is indeed  
the case.  
 
Let $\Phi$ be a truly marginal twisted sector operator. We would  
like to compute $g_{\rm cl}$ along the moduli space labeled by  
$\lambda$, the coefficient of $\Phi$ in the worldsheet Lagrangian, 
\eqn\pertmod{\delta\CL=\lambda\Phi(z,\bar z).} 
To this end we would like to repeat the derivation of  
\gengee\ for finite $\lambda$.  
As a first step, we will establish that \modtransfz\ 
holds for all $\lambda$. In particular, the coefficient of  
$\ZZ'$ on the r.h.s. is independent of $\lambda$. To show that, 
consider a general correlation function of primary operators 
on the torus, in the sector labeled by $g$ and $h$, as in \zzoorr, 
\eqn\corfnctor{ 
\langle\Phi_1(z_1)\Phi_2(z_2)\cdots\Phi_n(z_n)\rangle_{g,h;\tau}~.} 
Under the modular transformation $\tau\to -1/\tau$,  
$z_j\to -z_j/\tau$, this correlation function transforms as follows: 
\eqn\corfncttrans{\eqalign{ 
&\langle\Phi_1(-z_1/\tau)\Phi_2(-z_2/\tau)\cdots 
\Phi_n(-z_n/\tau)\rangle_{g,h;-1/\tau}=\cr 
&|\tau|^{2(h_1+h_2+\cdots+h_n)} 
\langle\Phi_1(z_1)\Phi_2(z_2)\cdots\Phi_n(z_n)\rangle_{h,g;\tau}\cr 
}} 
where $h_j$ are the scaling dimensions of $\Phi_j$ (which are assumed 
to be left-right symmetric). In particular, if the $\Phi_j$ all have 
dimension $h_j=\bar h_j=1$, and one integrates \corfncttrans\  
$\int d^2z_j$ over the torus, one finds that to all orders in  
$\lambda$ \pertmod, equation \modtransfz\ is still valid.  
 
In order to show that $g_{\rm cl}$ is independent of $\lambda$, 
one has to further prove that the limit as $\tau_2\to\infty$ of 
the partition sum on the r.h.s. of \modtransfz\ has the property 
that (in the sector with $h=1$), the coefficient of  
$(q\bar q)^{-{c\over24}}$ is 
independent of $\lambda$. Indeed, the derivative of this coefficient 
w.r.t. $\lambda$ (for generic $\lambda$) is proportional to 
the vacuum expectation value $\langle 0|\Phi(z,\bar z)|0\rangle$, which 
vanishes by conformal invariance.%
\foot{For systems with a continuous spectrum of scaling dimensions, 
the last statement is in general subtle, since expectation values  
of $\Phi(z,\bar z)$ in states with arbitrarily small scaling dimensions  
can in fact contribute. Here it is important that  
the partition sum on the r.h.s. of \modtransfz\ has $g\not=1$, so that 
at $\lambda=0$ the spectrum of scaling dimensions that contribute 
to it is discrete; the spectrum must remain discrete for all finite 
$\lambda$.} Therefore, we conclude that  
$g_{\rm cl}$ is independent of twisted moduli.   
 
We end this section with a few comments. 
As in the open string case, \nnn\ is expected   
to hold when we only deform the CFT by twisted sector operators  
(the analogs of open strings). Deformations by untwisted operators   
(which are analogs of closed strings) may increase or decrease   
$g_{\rm cl}$ (and will in general change the central charge   
as well, as discussed above). Also, there are subtleties which are    
familiar from the open string case, when some of the states    
contributing to the trace \ztwist\ are spacetime fermions    
(which contribute with a negative sign to the torus partition sum),    
which will be discussed below.  
  
We have phrased the discussion in the language of  
non-compact orbifolds of the form \aaa, but it should  
apply for more general non-compact backgrounds of string  
theory in which there are normalizable (localized)  
and delta-function normalizable (delocalized) excitations 
which satisfy \untwope.   
The discussion above should be applicable to the condensation of   
normalizable tachyons. Since they are localized, they presumably   
do not change the central charge $c$, but it is natural to   
conjecture that the high energy density of normalizable states   
decreases along RG flows. We will discuss some examples of this  
below. 
  
In the rest of the paper we will discuss a few classes of examples  
of the flows described in this section. We will verify the validity  
of \nnn\ and introduce some additional tools for studying such flows.


\newsec{$\IC/\IZ_n$ flows}  
  
Our first class of examples corresponds to $d=8$ and $\Gamma=\IZ_n$ in  
\aaa, \ie\ string theory on%
\foot{For an early discussion of string theory on $\IC/\IZ_n$,
see \DabholkarAI.}
\eqn\czn{\IR^{7,1}\times \IC/\IZ_n\ .}  
The chiral GSO projection of type II string theory acts  
non-trivially on the $\IC/\IZ_n$ factor, eliminating  
tachyons in the untwisted sector. It will be useful for  
our purposes to start with the non-chiral theory, with a  
diagonal GSO projection. One can think of this as type 0  
string theory on \czn. We will later discuss the modifications  
associated with passing from type 0 to type II.    
 
Note that the structure of the operator algebra \untwope\  
guarantees that classically (\ie\ on the sphere) the  
delocalized untwisted sector tachyon of type 0 remains  
unexcited along RG flows associated with condensing only  
twisted sector tachyons in infinite volume, and therefore  
it can be ignored.\foot{Just as in the open string case, one 
expects that it cannot be ignored when loop corrections are 
taken into account.} From a more general RG perspective,  
the statement is that there exist RG trajectories in which 
the untwisted tachyon is fine tuned to vanish. 
  
We will parametrize the complex plane $\IC$ in \czn\ by the  
worldsheet chiral superfield (we are working in the NSR formalism)  
\eqn\chirsup{\Phi=Z+\theta\psi+\bar\theta\bar\psi+\cdots\quad .}  
The orbifold action on the worldsheet fields is   
\eqn\orbact{\Phi\to \omega^j\Phi}  
where $ \omega=e^{2\pi i\over n}$.  
The theory contains $n-1$ twisted sectors, labeled by  
$j=1,2,\cdots, n-1$. To describe the ground states in the  
different sectors it is convenient to bosonize the fermions,  
\eqn\bosfer{\psi=e^{iH};\;\;\bar\psi=e^{i\bar H}\ .}  
An interesting set of operators in the orbifold CFT is  
\eqn\defxj{  
  X_j=\sigma_{j/n}\;\exp\left[i(j/n)(H-\bar H)\right]  
	\quad ;\qquad j=1,2,\cdots, n-1  
}  
where $\sigma_{j/n}$ is the bosonic twist $j$ operator \DixonQV.  
The operator $X_j$ has worldsheet scaling dimension   
\eqn\scdimxj{\Delta_j={j\over2n}\ .}  
In particular, the operators \defxj\ give rise to tachyons   
in the spacetime theory, with masses  
\eqn\massxj{{\alpha'\over4} M_j^2=\Delta_j-{1\over2}=  
-{1\over2}(1-{j\over n})\ .}  
It is also going to be useful to observe that CFT on $\IC/\IZ_n$  
is $N=(2,2)$ superconformal, with the R-charge generator  
$J=\psi\psi^*=i\partial H$ (and similarly for the right-movers).  
The operators \defxj\ have R-charge   
\eqn\rxj{R_j={j\over n}\ .}  
Comparing \scdimxj\ and \rxj\ we see that $X_j$ are chiral operators,  
a fact which will be useful below.   
  
It is not difficult to see that the ground states in all twist sectors  
are in fact of the form $X_j$, or $X_{n-j}^*$. Thus, in order to  
study the condensation of the most tachyonic state in each twisted  
sector, it is enough to study the $X_j$. This is the problem  
that will be analyzed here. There {\it are}  
excited tachyons in the spectrum which are not of the form \defxj.  
Including them in the analysis below is complicated, and we will  
not discuss them further.   
  
As discussed in section 2, in order to study the condensation of   
the twisted tachyons $X_j$, we would like to add the vertex operators    
\defxj\ to the worldsheet action\foot{More precisely, \defxj\    
is the bottom component of an N=2 chiral superfield.   
The worldsheet Lagrangian is perturbed by adding the top    
component of the superfield.  
For simplicity, we will also denote by $X_j$ the superfield  
whose lowest component is \defxj.}  
and  study the resulting RG flow.   
Since the operators \defxj\ are chiral, the    
corresponding deformation of the worldsheet Lagrangian is by an    
$N=2$ superpotential,   
\eqn\deformsup{\delta\CL=\lambda^j\int d^2\theta X_j\; +\;{\rm c.c.}}  
The $\lambda^j$ are $n-1$ complex couplings. The perturbation  
\deformsup\ breaks both conformal invariance and $U(1)_R$ (the trace  
of the stress tensor $\Theta$ \eee\ and the divergence of $J$   
are in the same multiplet under $N=2$ supersymmetry). Indeed,   
\scdimxj, \rxj, \deformsup\ imply that $\lambda_j$ carries   
R-charge $R(\lambda_j)=1-(j/n)$, and scaling dimension   
$\Delta(\lambda_j)=\half(1-(j/n))$. However, as is clear from   
\deformsup, the perturbation preserves $N=2$ supersymmetry for all   
$\lambda_j$. We would like to use this fact to obtain   
information about the possible RG flows in the model.

A strong constraint on the RG flow corresponding to \deformsup\  
comes from analyzing the chiral ring of the theory.   
Consider first the chiral ring of the undeformed theory,  
with all $\lambda_j=0$. It has two generators,  
\eqn\genchir{\eqalign{  
X=&X_1=\sigma_{1/n}\;\exp\left[(i/n)(H-\bar H)\right]\cr  
Y=&{1\over V_2}\psi\bar\psi={1\over V_2}\;\exp\left[i(H-\bar H)\right]\ .  
}}  
Note that $Y$ is the volume form of $\IC/\IZ_n$   
divided by the volume of the cone, $V_2$ 
(\eg\ regularized as $S^2/\IZ_n$ \CecottiTH).  
The higher chiral twisted operators $X_j$ can   
be thought of as powers of $X$, $X_j=X^j$, and there is a relation   
\eqn\ringrel{X^n=Y\ .}  
One can in fact focus on the chiral ring while eliminating the  
rest of the dynamics, by performing a topological twist. The details  
of this are not important for us here, and we will not discuss them   
further (see \CecottiTH\ for a discussion). 
  
When one turns on the superpotential \deformsup, the chiral ring  
gets deformed. In particular, the relation \ringrel\ is modified to  
\eqn\defring{X^n+\sum_{j=1}^{n-1} g_j(\lambda) X^j=Y}  
where $g_j(\lambda)$ are polynomials in the couplings $\lambda$  
(but by holomorphy, are independent of $\lambda^*$), with   
R-charge $R(g_j)=1-(j/n)$. To leading order one has  
$g_j=c_j\lambda_j +O(\lambda^2)$, where $c_j$ are certain  
non-vanishing constants. The fact that there are higher  
order corrections to this relation is a standard feature  
that has to do with coordinate choices on the space of couplings.   
This too is not of interest to us here, and we will ignore the   
difference between $g_j$ and $\lambda_j$.   
  
The flow to the infrared can now be analyzed much like 
similar flows in  
Landau-Ginzburg models \refs{\ZamolodchikovDB,\KastorEF}.  
If one tunes $g_1, g_2,\cdots, g_{n'-1}$ to zero, and the  
leading non-vanishing relevant coupling is $g_{n'}$, \defring\  
describes a flow from a theory with ring relation $X^n=Y$  
in the UV to one corresponding to   
\eqn\gnprime{g_{n'}X^{n'}=Y}  
in the infrared.  Equation \gnprime\ looks like the chiral ring relation   
corresponding to the orbifold $\IC/\IZ_{n'}$. We conclude that the  
deformed ring \defring\ describes a sequence of transitions  
from $\IC/\IZ_n$ to other orbifolds of the same form but with lower  
$n$. Turning on generic $g_j$ resolves the singularity completely.  
This is consistent with the conclusions of \AdamsSV.  
  
It should be mentioned that the above arguments do not determine  
the IR limit uniquely, since they only fix the chiral ring of the  
IR theory, and it is possible that two different CFT's share the  
same chiral ring. Nevertheless, the structure of the ring\foot{By using   
the results of \CecottiTH, one can obtain additional information   
about the deformed chiral ring, such as the structure constants.   
We will not pursue this here.} places  
a severe constraint on the possible endpoints of the flow, and it  
is very likely that the above interpretation of the IR fixed point  
is correct.   
  
Following the discussion of section 2, it is of interest to compute  
$g_{\rm cl}$ for the $\IC/\IZ_n$ orbifolds and check whether it is   
an increasing function of $n$. Thus, we would like to compute  
\eqn\ztwtan{  
  \ZZ_{\rm tw}(\tau,\bar\tau)=\sum_{s=1}^{n-1}  
	{\rm Tr}_s q^{L_0-{c\over 24}} \bar q^{\bar L_0-{c\over 24}}\ ,}  
where the trace runs over $\IZ_n$ invariant  
states in the $s$-twisted sector.   
Using standard techniques one can show that the   
twisted partition sum is given by   
\eqn\integer{\ZZ_{\rm tw}(\tau,\bar\tau)=  
\sum_{s=1}^{n-1}{1\over 2n} \sum_{t=0}^{n-1}  
\sum_{\epsilon_1, \epsilon_2=0,\half}  
\biggl\vert {\vt{\epsilon_1 + s/n}{\epsilon_2+t/n}{0}  
\over  
\vt{\half -  s/n}{\half +t/n}{0} }\biggr\vert^2  
\ .}  
Performing the modular transformation $\tau\to-1/\tau$,   
and comparing to the definition of $g_{\rm cl}$ \mmm, \zgcl,  
one finds  
\eqn\asympt{  
g_{\rm cl}(n)={1\over n}  
\sum_{s=1}^{n-1} {1\over (2 \sin\pi s/n)^2}  
\ ,}  
in agreement with \gengee. 
The sum in \asympt\ is evaluated in appendix A. One finds  
\eqn\gclfinal{  
  g_{\rm cl}(n)={1\over 12}(n-{1\over n})  
}  
which is indeed a monotonically increasing function of $n$, in  
agreement with the ``$g_{\rm cl}$-conjecture'' \nnn. This provides  
further support for the flows proposed above.   
Note also that \gclfinal\ vanishes for $n=1$, in accord with the  
intuition that for $\IC/\IZ_1=\IC$, there are no localized states.   
  
The discussion above actually needs to be refined somewhat, in a way  
familiar from the study of Landau-Ginzburg theories.  Imagine tuning  
the couplings in \defring\ in such a way that the relation \ringrel\  
is deformed to  
\eqn\defnonek{(X-x_1)^{n_1}(X-x_2)^{n_2}\cdots(X-x_k)^{n_k}=Y}  
with $n=\sum_{i=1}^k n_i$. The $x_i$ are functions of the couplings  
$g_j$. When we flow to the IR they grow, and the system is expected  
to split into decoupled lower rank singularities. In \gnprime\ we   
focused on a particular vacuum (at $X=0$), but in general there are   
additional vacua corresponding to $X$ near other $x_i$. Since the 
Hilbert space of localized states of the extreme IR theory is a
direct sum of the Hilbert spaces corresponding to $\IC/\IZ_{n_i}$,   
we have  
\eqn\guvirsum{\eqalign{  
g_{\rm \sst UV}=&{1\over 12}(n-{1\over n})\cr   
g_{\rm \sst IR}=&{1\over 12}\sum_{i=1}^k(n_i-{1\over n_i})\ .}}  
One can check that the $g_{\rm cl}$-conjecture \nnn\ is satisfied:  
\eqn\checkgconj{n-{1\over n}>\sum_i(n_i-{1\over n_i})}  
  
Qualitatively, the transition  
in \guvirsum\ can be thought of as a process in which a $\IC/\IZ_n$  
cone splits into $\IC/\IZ_{n_i}$ cones that are decoupled from each   
other (see figure 1).

\bigskip  
{\vbox{{\epsfxsize=4in  
        \nobreak  
    \centerline{\epsfbox{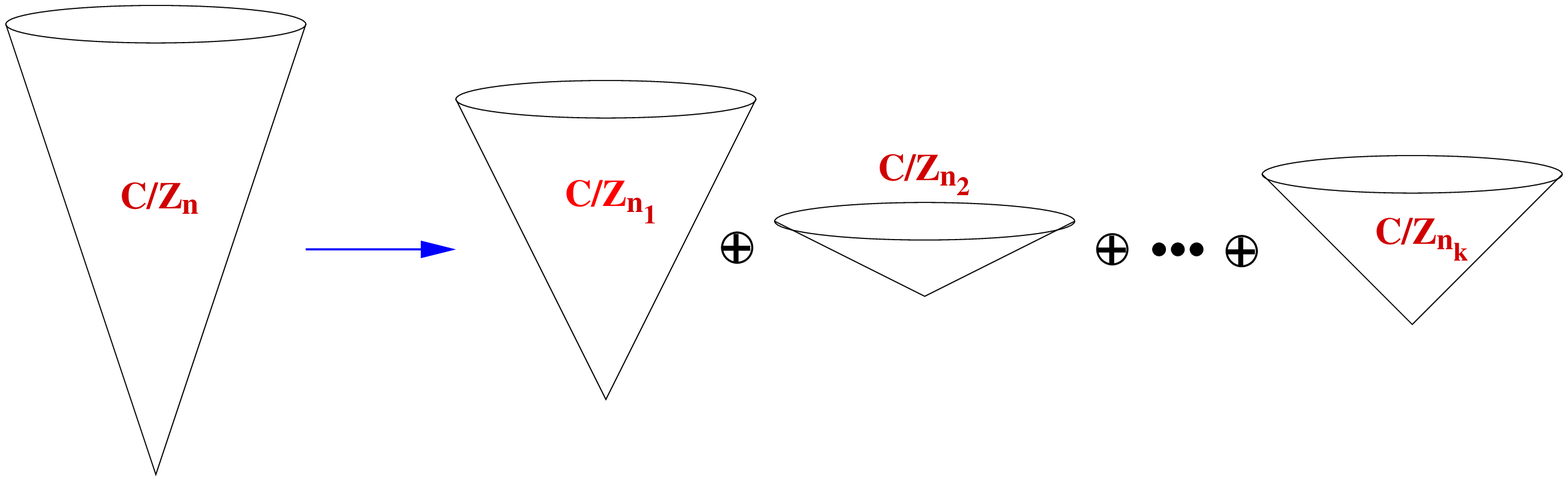}}  
        \nobreak\bigskip  
    {\raggedright\it \vbox{  
{\bf Figure 1.}  
{\it   
Tachyon condensation of the form in equation \defnonek\  
leads to a set of decoupled cones. 
} }}}}  
    \bigskip} 
\noindent  
It is natural to ask how the process of ``cone splitting'' 
in figure 1 occurs in real time. There are two complications 
with answering this question using the discussion above. First, 
we have analyzed the RG evolution of the system, which is an 
off-shell process in the spacetime theory. One might hope that 
the time evolution is similar, but this is not known to be the   
case in general (\eg\ here time evolution breaks worldsheet  
$N=2$ supersymmetry, while RG evolution preserves it). 
 
Second, the splitting 
of cones depicted in figure 1 is naturally described in the 
coordinate $X$, which is the field space of the twist field  
\genchir. This coordinate can be roughly thought of as associated 
with winding around the tip of the cone parametrized by $Z$ \chirsup.  
All the relevant perturbations respect the $U(1)$ rotation symmetry  
around the tip, as can be checked by computing the action of the  
rotation generators on the twist fields.  
 
Since $X$ is essentially the vertex operator of the twisted tachyon   
field (see \genchir), one can think of it as a (complex) dimension  
of field space in the target space theory. Figure 1 seems to suggest  
that it is natural to think of it as an extra complex dimension in  
spacetime. The fragmentation of  cones in figure 1 happens in this   
dimension, in a way compatible with the $U(1)$ rotational 
symmetry in $Z$. 
 
So far we have been discussing the diagonal $\IC/\IZ_n$ theory,   
or type 0 string theory on \czn. We would now like to discuss   
the generalization to the chirally GSO projected theory, or type II  
on \czn.  
The latter can be obtained by gauging a chiral $\IZ_2$ symmetry  
in the diagonal (type 0) theory. In flat spacetime, $\IR^{9,1}$,  
one gets type II from type 0 by gauging the chiral $\IZ_2$ symmetry  
$(-)^{F_L}$. In the background \czn, the symmetry in question acts  
as $(-)^{F_L}$ on $\psi_0,\psi_1,\cdots,\psi_7$, and as  
\eqn\hnpi{H\to H+n\pi;\;\; \bar H\to \bar H}   
on $\psi, \bar\psi$ (see \bosfer).   
  
In order for \hnpi\ to act as $(-)^{F_L}$ in the untwisted  
sector (and thus eliminate untwisted sector tachyons),   
one must take $n$ to be odd \AdamsSV.  
In the twisted sectors, \hnpi\ takes  
\eqn\xjtrans{X_j\to(-)^j X_j}  
(see \defxj), and since the $(-1,-1)$ picture vertex operators  
must be odd under chiral GSO, only the states with odd $j$  
survive. In the sectors with even $j$, $X_j$ is projected out,  
and so the lowest lying state is  
\eqn\tildexj{\tilde X_j=  
\sigma_{j/n}\;\exp\left[i(\coeff jn-1)(H-\bar H)\right]\ ,}  
which is nothing but the complex conjugate of $X_{n-j}$,  
\eqn\xxstarj{\tilde X_j=X_{n-j}^*\ .}  
As a check, the mass of the lowest lying states for $j\in 2\IZ+1$  
is given by \massxj, while for $j\in 2\IZ$ \xxstarj\ implies that  
\eqn\masseven{{\alpha'\over4} M_j^2=-\half{j\over n}\ ,}  
in agreement with the light-cone Green-Schwarz analysis of \AdamsSV.  
  
We see that the effect of the chiral GSO projection is to restrict  
to $n\in 2\IZ+1$, and allow only couplings $\lambda_j$,  
$g_j$ with odd $j$ in \deformsup, \defring.   
  
The analysis of flows is similar to the one for the diagonal  
theory, with a few interesting differences. Since $n,j\in 2\IZ+1$  
in \defring, the analog of the deformed relation \defnonek\ is now:  
\eqn\xgsoch{  
X^{n'}(X-x_1)^{n_1}(X+x_1)^{n_1} (X-x_2)^{n_2}(X+x_2)^{n_2}\cdots  
=Y\ }  
with $n=n'+2\sum_i n_i$, $n'$ an odd integer, and $x_i\not=0$.  
The vacuum structure is now more intricate than in the type 0 case.  
The vacuum at $X=0$ corresponds to a $\IC/\IZ_{n'}$ singularity in  
type II string theory. This is similar to what happens in the diagonal  
theory. Here, because of chiral GSO, only flows between odd $n$'s  
occur. The vacua at $X=\pm x_i$ are more  
interesting, since they break the $\IZ_2$ symmetry $X\to -X$. Thus,  
chiral GSO \xjtrans, which is a discrete gauge symmetry, is  
{\it spontaneously broken} in these vacua. Hence, the infrared  
theory at vacua like $X=x_1$ in \xgsoch\ is type 0 string theory  
on $\IC/\IZ_{n_1}$.  
  
We conclude that type II string theory on $\IR^{7,1}\times \IC/\IZ_n$  
is connected via tachyon condensation to a decoupled sum of a type  
II string theory on a lower singularity, $\IC/\IZ_{n'}$, with $n'<n$  
and odd, and a set of type 0 vacua on $\IC/\IZ_{n_i}$, $i=1,\cdots, k$.  
To check whether the $g_{\rm cl}$-conjecture is satisfied, we note  
that:  
\item{(1)} It is at first sight not clear how to define  
$g_{\rm cl}$ in a theory with both (spacetime) bosons and fermions.  
We propose to define it via the density of bosonic states \KutasovPF,  
as in open superstring theory \HarveyGQ.   
\item{(2)} The value of $g_{\rm cl}$ for the type II theory on \czn\  
is one half of that for the diagonal (type 0) theory \gclfinal, since  
half of the states in the (NS,NS) and (R,R) sectors survive the   
projection. More formally, this factor of two comes about   
because in the diagonal theory we insert the projector  
\eqn\diaggso{  
  \hf (1 + (-1)^{F_L + F_R})  
}  
and both terms contribute, while in the type II theory we  
project with  
\eqn\chigso{  
  \coeff14 (1 + (-1)^{F_L } + (-1)^{F_R} + (-1)^{F_L + F_R})  
}  
and only the first and fourth terms contribute  
to the asymptotic density of states.  
\item{(3)} $X=x_i$ and $X=-x_i$ in \xgsoch\ do not give independent  
vacua, since they are related by a gauge symmetry (the broken chiral  
GSO).  
  
\noindent  
Hence, the $g_{\rm cl}$-conjecture is in this case the statement  
that   
\eqn\gtypetwo{\half(n-{1\over n})>\half(n'-{1\over n'})+  
\sum_{i=1}^k(n_i-{1\over n_i})\ ,}  
which is indeed valid, since $n=n'+2\sum_i n_i$ (see \xgsoch).  
  
The type 0 components of the infrared theory have  
delocalized tachyons, and thus will presumably disappear when   
these condense. There is always a  
type II component in the background (corresponding to the  
vacuum at $X=0$), which does not have delocalized tachyons,  
and in it the physics is similar to that described above:  
the deficit angle of the cone $\IC/\IZ_n$ decreases along the  
flows, and eventually it decays to type II string theory on  
a large smooth space, $\IR^{9,1}$.  
  
Another approach to the study of twisted tachyon condensation, 
which was discussed recently in \AdamsSV, is to analyze the 
dynamics of probe D0-branes on the orbifold $\IC/\IZ_n$. At 
low energies this is described by a quiver gauge theory with 
gauge group $U(1)^n$ and matter given in \AdamsSV. The Higgs 
branch of the moduli space of vacua of this theory is the orbifold 
itself. The twisted closed string tachyon v.e.v.'s $\lambda_j$ 
\deformsup\ give rise to parameters in the open string Lagrangian. 
 
In principle, if one knows the precise form of the D-brane Lagrangian 
as a function of $\{\lambda_j\}$, one can analyze the moduli space 
and deduce from it the geometry of the orbifold after tachyon 
condensation. In practice, it is difficult to determine the $\lambda$ 
dependence of the D-brane Lagrangian for all $\lambda$. For small 
$\lambda$, one can show that the leading effect of tachyon condensation 
is to turn on (analogs of) Fayet-Iliopoulos D-terms for the $U(1)^n$ 
gauge theory, and to modify the gauge couplings.  
In appendix B we review the computation of \DouglasSW\ and show that  
the FI parameters coupling to the D-terms   
 $\{\zeta_j\}$ are given to first order in $\lambda$ by 
\eqn\xijlam{\zeta_j= C\sum_{j'=1\atop j'\in 2{\bf Z}+1}^n   
{\rm Im} \biggl( e^{-2\pi i jj'/n} \lambda_{j'}\biggr) \ ,}   
where $C$ is a real constant.  The gauge couplings $1/g_j^2$ are  
modified as follows: 
\eqn\gaugelam{\eqalign{   
{1\over g_j^2}=&{1\over g_0^2}(1+B_j)\cr   
B_j=&C'\sum_{j'=1\atop j'\in 2{\bf Z}+1}^n{\rm Re} 
 \biggl( e^{-2\pi i jj'/n} \lambda_{j'}\biggr)\cr   
}}   
with $C'$ a real constant. By analyzing the Higgs branch of the  
theory with the couplings \xijlam, \gaugelam\ turned on, one may  
hope to get an indication of the geometry after tachyon condensation. 
Of course, at small $\lambda$, one expects at best to get a small 
patch of the new geometry near the (deformed) tip of the cone. 
 
The leading, small $\lambda$, effect of tachyon condensation is  
to break spontaneously the quiver gauge group $U(1)^n$, due to 
the ``D-term potential'' associated with $\{\zeta_j\}$ \xijlam. 
For generic $\{\lambda_j\}$ (\ie\ generic $\{\zeta_j\}$ subject 
to the constraint $\sum_j\zeta_j=0$), one finds \AdamsSV\ that 
the gauge group is completely broken, and the moduli space is a 
smooth non-singular manifold. This is consistent with our results, 
since for generic $\{\lambda_j\}$, one has $n'=1$ in \xgsoch, and 
the extreme IR limit of the theory is type II string theory on 
$\IR^{7,1}\times \IC$, together with a number of  
decoupled unstable type 0 vacua. The latter do not seem  
to be visible in the quiver analysis at small $\lambda$.  
 
As discussed in \AdamsSV, one can fine tune the $\{\zeta_j\}$ \xijlam\ 
such that $U(1)^n$ is broken to $U(1)^{n'}$ with any $n'<n$. For 
$n'$ odd one finds the quiver for $\IC/\IZ_{n'}$, and it is natural 
to interpret this as a small $\lambda$ indication of the flow described 
by \xgsoch. Our results in fact show that in this 
case the higher order corrections in $\lambda$ do not modify the 
leading order predictions. 
 
For $n'$ even the situation is different. Here there is no candidate 
$\IC/\IZ_{n'}$ CFT to serve as the endpoint of the flow, and indeed the 
probe gauge theory is not of quiver form. Thus, there are two possibilities. 
One is that there is in fact a consistent background describing type II 
string propagation (without bulk tachyons) on $\IC/\IZ_{n'}$  with  
$n'\in 2\IZ$, but that this background cannot be thought of as an orbifold 
in the usual free field theory sense. Such backgrounds were referred to as 
``quasi-orbifolds'' in \AdamsSV. This possibility is not supported by our 
analysis, which shows no evidence of exotic CFT's at the endpoint of tachyon 
condensation. 
 
A second possibility is that the flow $\IC/\IZ_n\to \IC/\IZ_{n'}$ indicated 
by the leading order analysis is qualitatively modified by higher order 
effects in $\lambda$. This possibility is consistent with our results. 
 
There are two kinds of effects that are known to be important here  
(there  may be other effects as well).  
One is higher order corrections to the   
potential for the quiver fields. This could change qualitative features   
such as the rank of the maximal unbroken gauge group, and thus the   
singularity structure of the orbifold after tachyon condensation.   
For example, it could be that the exact potential for the fields on   
the D-brane does not have any vacua in which $U(1)^n$ is broken to   
$U(1)^{n-1}$, for any choice of $\{\zeta_j\}$, unlike the leading order   
potential studied in \AdamsSV. 
 
Another important effect is related to the behavior of the gauge   
couplings on the quiver. Naively, one might expect that since the   
D-brane probe is studied in the weak coupling limit, $g_s\to 0$,   
the only way factors of the quiver can disappear is by the 
Higgs mechanism,  as described in detail in \AdamsSV.  
But, one of the lessons of the   
work on open string tachyon condensation on non-BPS branes in the  
last few years is that there is another, stringy, way for D-branes   
to disappear, by a process which formally looks like classical   
confinement.\foot{There has been some debate as to the precise   
mechanism by which the D-branes disappear   
\refs{\SenMD,\YiHD,\BergmanXF,\HarveyNA,\KlebanPF}; this is   
irrelevant for the present discussion.}  In this process, the   
whole action on the D-brane goes to zero (see \eg\ \KutasovQP).   
 
It is  possibile that twisted closed string tachyon 
condensation  leads to similar effects on the probe D-brane. In  
order to analyze this, one would have to compute the D-brane action 
as a function of $\{\lambda_j\}$, and check whether it goes 
to zero when $\{\lambda_j\}$ approach their IR fixed point values. This 
would involve computing the gauge couplings \gaugelam, the tensions of 
the fractional branes, and other quantities, to all orders in $\lambda$. 
This is beyond the scope of our analysis, but it is interesting that  
to first order in $\{\lambda_j\}$, at least one of the couplings  
$\{g_j\}$ in \gaugelam\ 
is getting stronger (since $\sum_j B_j=0$). We believe that the 
``classical confinement'' mechanism is important in this problem. 
 
Finally, one can ask what happens  
when one allows $N=2$ supersymmetric breaking 
on the worldsheet. One might hope that this allows independent control 
over all the low energy parameters of the quiver, since there are of order 
$n^2$ twisted sector tachyons. However, these couplings are not free  
parameters -- the tachyon v.e.v.'s in the IR are solutions of the closed  
string beta function equations, which are generically isolated points  
in the space of couplings. The quiver gauge couplings are functions  
of the tachyon v.e.v.'s and take some particular values at the IR fixed  
points. There is a good chance that some of the gauge couplings  
are shifted substantially toward strong coupling during the flow,  
just as in the N=2 preserving examples. More control over the quiver  
analysis  (or some other way of analyzing N=1 flows) is needed before  
one can claim that ``quasi-orbifold'' solutions to the string equations  
of motion exist.

  
  
\newsec{$\IC^2/\IZ_n$ flows}  
 
In section 3 we saw that $\IC/\IZ_n$ orbifolds exhibit an interesting 
pattern of infrared instabilities and decays. Higher dimensional 
noncompact orbifolds $\IC^m/\Gamma$ with $m>1$ are expected to  
exhibit an even richer behavior. As a first step towards their study,  
in this section we discuss, following \AdamsSV, a class of  
$\IC^2/\IZ_n$ orbifolds where $\IZ_n$ acts as follows: 
\eqn\znp{ 
  \RR(Z_1, Z_2)=(\omega Z_1,\omega^{p}Z_2) 
	\quad,\qquad p\in\IZ\ ,  
}  
$\omega=\exp[2\pi i/n]$, and $0<|p|<n$, with  
$p,n$ relatively prime.  We will denote the  
orbifold \znp\ by $\IC^2/\IZ_{n(p)}$, as in \AdamsSV.  
Quotients of the type \znp\ are known as Hirzebruch-Jung 
singularities, and their geometry is discussed for instance 
in \refs{\bpv,\fulton}.  
 
The orbifolds $\IC^2/\IZ_{n(p)}$ and $\IC^2/\IZ_{n(-p)}$  
are related by a change of complex structure, $Z_2\to Z_2^*$, 
and are thus isomorphic. We will find it convenient to keep 
the complex structure fixed, and discuss the cases of positive 
and negative $p$ separately. The Kahler form is  
\eqn\kahler{K=dZ_1\wedge d Z_1^*+dZ_2\wedge d Z_2^*} 
The worldsheet conformal field theory has a corresponding 
left-moving $U(1)_R$ current  
\eqn\uoner{J(z)=\psi_1\psi_1^*+\psi_2\psi_2^*} 
and similarly for the right-movers. 
 
As for $\IC/\IZ_n$, the chiral ring of the orbifold CFT will play 
an important role in our analysis. The building blocks out of which 
chiral operators in $\IC^2/\IZ_{n(p)}$ are constructed are the twist  
fields $X^{\sst(1)}_j$, $X^{\sst(2)}_j$ $\defxj$ for the two complex  
planes parametrized by $Z^1$, $Z^2$, and the corresponding 
volume forms $Y^{\sst(1)}$, $Y^{\sst(2)}$ \genchir. Combining the two, 
one finds the twisted sector chiral operators 
\eqn\znptwist{X_j=X^{\sst(1)}_j X^{\sst(2)}_{n\{\!\frac{jp}n\!\}}} 
where $j=1,2,\cdots, n-1$ labels the twisted sectors, and 
$\{x\}$ is the fractional part of $x$, $\{x\}=x-[x]$, with  
$[x]$ the integer part of $x$ (the largest integer $\le x$). 
Note that by definition $0\le \{x\}<1$.  
The R-charges of the chiral operators \znptwist\ are 
\eqn\rchtwist{R_j={j\over n}+ \left\{{jp\over n}\right\}.} 
 
For type II strings, one has to  perform further a chiral GSO 
projection, which can be described in a way analogous to the 
$\IC/\IZ_n$ case (see \hnpi). It acts on the bosonized fermions 
$H_1$, $H_2$ (which are defined by $\psi_j=\exp(iH_j)$, as in \bosfer)  
as the chiral $\IZ_2$ shift  
\eqn\ctwogso{ 
  H_1\rightarrow H_1+p\pi\quad,\qquad H_2\rightarrow H_2-\pi\ . 
} 
In the untwisted sector of the orbifold, \ctwogso\ must reduce  
to the standard $(-)^{F_L}$; this implies that $p$ must be odd,  
in agreement with \AdamsSV. In the twisted sectors, 
\ctwogso\ projects out operators with $[jp/n]\in 2\IZ$ and keeps 
those with $[jp/n]\in 2\IZ+1$. One can show that this definition of  
chiral GSO in the NSR framework gives the same spectrum as the  
light-cone Green-Schwarz analysis.  
 
\lref\DixonBG{ 
L.~J.~Dixon, 
``Some World Sheet Properties Of Superstring Compactifications,  
On Orbifolds And Otherwise,'' {\it Lectures given at the 1987  
ICTP Summer Workshop in High Energy Physics and Cosmology,  
Trieste, Italy, Jun 29 - Aug 7, 1987}. 
} 
 
Operators whose R-charge \rchtwist\ is smaller than one give rise 
to tachyons in spacetime. One new element in the discussion compared  
to the $\IC/\IZ_n$ case, is the appearance of massless states with 
flat potentials, corresponding to operators \znptwist\ with R-charge 
{\it equal} to one. As is well known \DixonBG, such operators correspond 
to exactly marginal operators on the worldsheet, and one can study the 
theory as a function of their coefficients in the action (moduli).   
As we saw in section 2, $g_{\rm cl}$ is independent of the moduli,  
for finite changes of moduli. If a modulus is taken strictly 
to infinity (in the natural metric on moduli space), $g_{\rm cl}$ can 
jump discontinuously (down), since in the limit some localized states 
might decouple.  
 
This behavior is familiar from the study of boundary CFT's in D-brane  
physics. As a simple example, consider a collection of $n$ D0-branes 
on an infinite line. For any finite separation of the D-branes (which 
corresponds to an expectation value of a massless open string field, 
or boundary modulus), $g_{\rm op}$ \bbb\ is independent of the  
separation, $g_{\rm op}=n^2 g_1$, where $g_1$ is the value for a single  
D0-brane. If we take a limit in which the collection of D-branes  
splits into two groups of $n_1$ and $n_2$ D-branes ($n=n_1+n_2$)  
separated by an infinite distance, $g_{\rm op}$ decreases to  
$g_{\rm op}=(n_1^2+n_2^2) g_1$, since the states associated with 
open strings stretched between the two clusters of branes become  
infinitely massive and decouple. One expects the same general  
behavior for $g_{\rm cl}$ as a function of twisted sector moduli.  
 
In this section we will discuss the physics associated with 
relevant and marginal perturbations of $\IC^2/\IZ_{n(p)}$  
by chiral operators of the form \znptwist. One of the main tools that 
we will use to study such flows is the classical geometry of  
Hirzebruch-Jung quotient singularities. As we describe in the next 
subsection, many properties of the $\IC^2/\IZ_{n(p)}$ CFT are 
directly related to the geometry of Hirzebruch-Jung singularities.  
This relation will allow us to develop a rather detailed picture  
of the flows in these systems.  
 
\lref\KutasovTE{ 
D.~Kutasov, 
``Orbifolds and Solitons,'' 
Phys.\ Lett.\ B {\bf 383}, 48 (1996) 
hep-th/9512145. 
}

For the special case $p=-1$, we will use another tool to study the  
system. In this class of orbifolds, the worldsheet symmetry is enhanced  
to $N=4$ superconformal symmetry, and the GSO projection \ctwogso\  
leads to a spacetime supersymmetric background. In this case one can  
use a duality of the orbifold CFT $\IC^2/\IZ_{n(-1)}$ to a system of  
NS5-branes \refs{\OoguriWJ,\KutasovTE}, which provides a nice  
geometric picture of the possible flows. This is described in section 
4.2.   
 
One of our main purposes below is to verify the validity of the  
conjecture \nnn\ for the different cases. For the class of  
orbifolds \znp, $\gcl$ is given by\foot{This holds for
 the type 0 case. For
type II there is an additional factor of $1/2$, as in section 3.} 
\eqn\gengctwo{  
  g(n,p) = {1\over n}  
	\sum_{s=1}^{n-1} {1\over [4 \sin(\pi s/n) \sin(\pi p s/n)]^2} \ .  
}  
Below we will examine in detail the cases 
$p=\pm1$, for which (see appendix A) 
\eqn\gnone{  
  \gcl(n,\pm1)= { (n^2+11) (n^2-1)\over 45\cdot 16 n} \ ; 
} 
and $p=\pm3$, for which 
\eqn\gnthree{ 
  \gcl(n,\pm3)= \cases{ 
	{(n^4 + 210 n^2 - 80 n - 291)\over 405 \cdot 16 n}   
		& $n = 2~{\rm mod}~3$\cr  
	& \cr 
 	{(n^4 + 210 n^2 + 80 n - 291)\over 405  \cdot 16 n}  
		& $n = 1~{\rm mod}~3$\ .}  
} 
For general $n$, $p$, the function \gengctwo\ is rather complicated 
(see Appendix A); it is useful to note that in the limit  
$n\to\infty$, $p$ fixed, it simplifies: 
$\gcl\simeq\coeff1{720} \frac{n^3}{p^2}$. 
 
\subsec{Chiral rings, Hirzebruch-Jung geometry,  
and singularity resolution} 
 
In this subsection we will briefly summarize the Hirzebruch-Jung 
theory of singularity resolution for cyclic surface 
singularities and its relation to the structure  
of the chiral ring of the corresponding orbifold $N=2$ SCFT. 
We will use some of the techniques of toric geometry without 
detailed explanation. Useful references on this material 
include \refs{\bpv,\fulton,\oda,\danilov} as well as the treatments 
aimed more towards physicists in  
\refs{\agm,\hkt,\MorrisonPlesser,\GreeneCY,\Skarke}. 
 
\subsubsec{~Hirzebruch-Jung geometry} 
 
The singular geometry corresponding to the quotient \znp\ is defined  
in toric geometry by a cone consisting of positive real linear 
combinations  of 
two generators,  
$v_f \equiv v_{r+1}=e_2$ and $v_i \equiv v_0=ne_1 -p e_2$ as shown  
in figure 2.  
 
\bigskip  
{\vbox{{\epsfxsize=2.5in  
        \nobreak  
    \centerline{\epsfbox{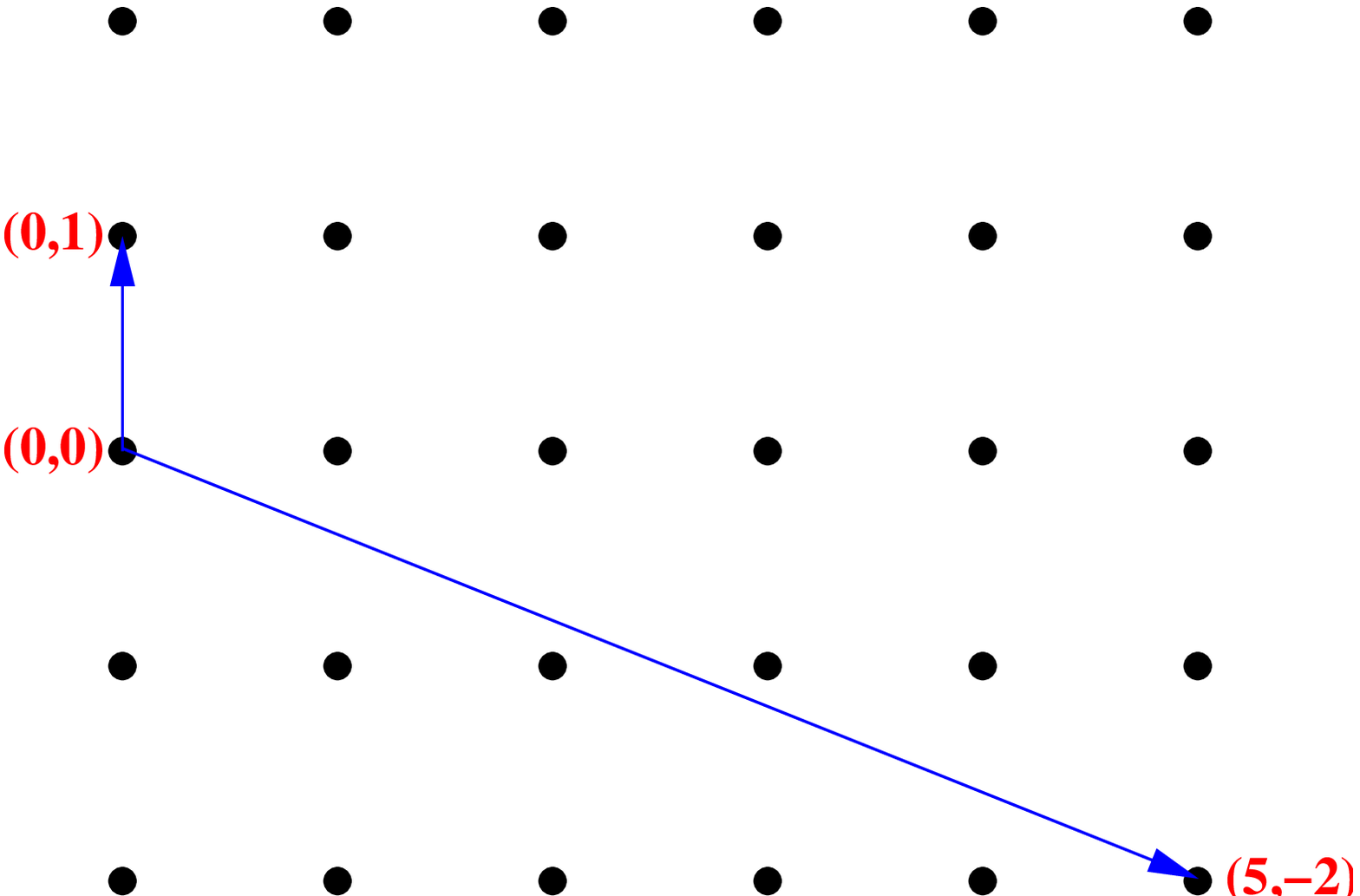}}  
        \nobreak\bigskip  
    {\raggedright\it \vbox{  
{\bf Figure 2.}  
{\it   
Cone defining the quotient singularity $\IC^2/\IZ_{n(p)}$\ 
for $(n,p)=(5,2)$. 
} }}}}  
    \bigskip}           
 
Here $e_1,e_2$ are a basis for the lattice $N=\IZ^2$. A  
resolution of singularities is given by adding a set of interior 
vectors $v_j$, $j=1, 2, \cdots r$ lying between $v_0$ and $v_{r+1}$ 
such that 
 
\item{1.} The $v_j$ are rational, meaning they lie in $N$. 
\item{2.} Each successive pair of vectors $(v_0,v_1),(v_1,v_2), \cdots 
(v_r,v_{r+1})$ spans the lattice $N$ as a $\IZ$-module. 
 
\noindent 
One implication of this structure is that there exist integers $a_j$ 
such that 
\eqn\arelns{a_j v_j = v_{j-1}+v_{j+1} } 
for all interior vectors $v_j$. 
One can show that each interior vector corresponds to 
an exceptional divisor $E_j \cong \IP^1$ with self-intersection 
number $-a_j$ and with $E_j$ intersecting $E_{j+1}$ once. 
 
We may summarize 
this in a resolution diagram: 
 
\bigskip  
{\vbox{{\epsfxsize=3in  
        \nobreak  
    \centerline{\epsfbox{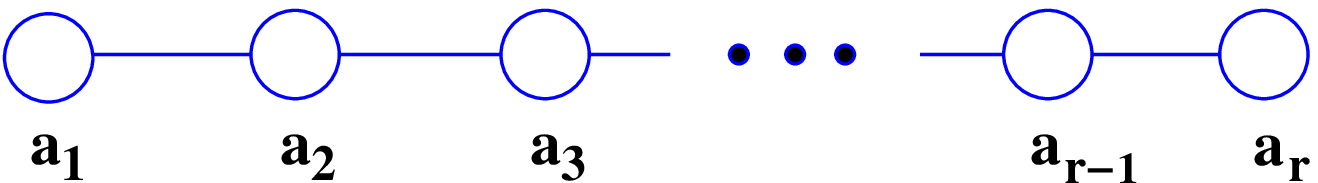}}  
        \nobreak\bigskip  
    {\raggedright\it \vbox{  
{\bf Figure 3.}  
{\it   
Resolution diagram of $\IC^2/\IZ_{n(p)}$.  The links denote  
unit intersection of successive curves in the resolution, 
and the labels on the nodes denote their self-intersection numbers. 
} }}}}  
    \bigskip}

The combinatorics of this construction are directly captured by 
the continued fraction 
expansion of $n/k$, where $k=p$ for $p>0$, and  
$k=n+p$ for $p<0$ (in other words, $0<k<n$ and $k=p$~mod~$n$). 
The continued fraction expansion is defined as 
\eqn\contfrac{ 
  \frac nk=~a_1-\frac{1}{~a_2-\frac{1}{~a_3-\ldots}} 
	~\equiv~ [a_1,a_2,a_3,\ldots,a_r]\qquad. 
} 
with integers $a_j \ge 2$. This is called the Hirzebruch-Jung continued 
fraction of $n/k$. There is a minimal resolution of the singularity  
where the $\{a_j\}$ in \arelns\ are the same as the 
integers in \contfrac. It is minimal in the sense that no 
curve can be blown down while leaving a nonsingular variety.

A familiar example is the resolution of singularities for the ALE 
space $\IC^2/Z_{n(-1)}$. In toric geometry this singularity is 
described by the cone generated by $v_0= ne_1 - (n-1)e_2$ and 
$v_n=e_2$ and corresponds to an $A_{n-1}$ singularity, 
$\IC[Y_1,Y_2,Y_3]/(Y_3^{n-1}=Y_1Y_2)$. The resolved geometry 
is described by a fan with $n-1$ interior vectors as illustrated 
in figure 4 for the case $n=4$. The resolved geometry has 
$n-1$ $\IP^1$'s which intersect according to the Dynkin diagram 
of $A_{n-1}$. This agrees with the 
continued fraction expansion  $n/(n-1) = [2^{n-1}]$, 
with powers denoting repeated entries. 
 
\bigskip  
{\vbox{{\epsfxsize=2.2in  
        \nobreak  
    \centerline{\epsfbox{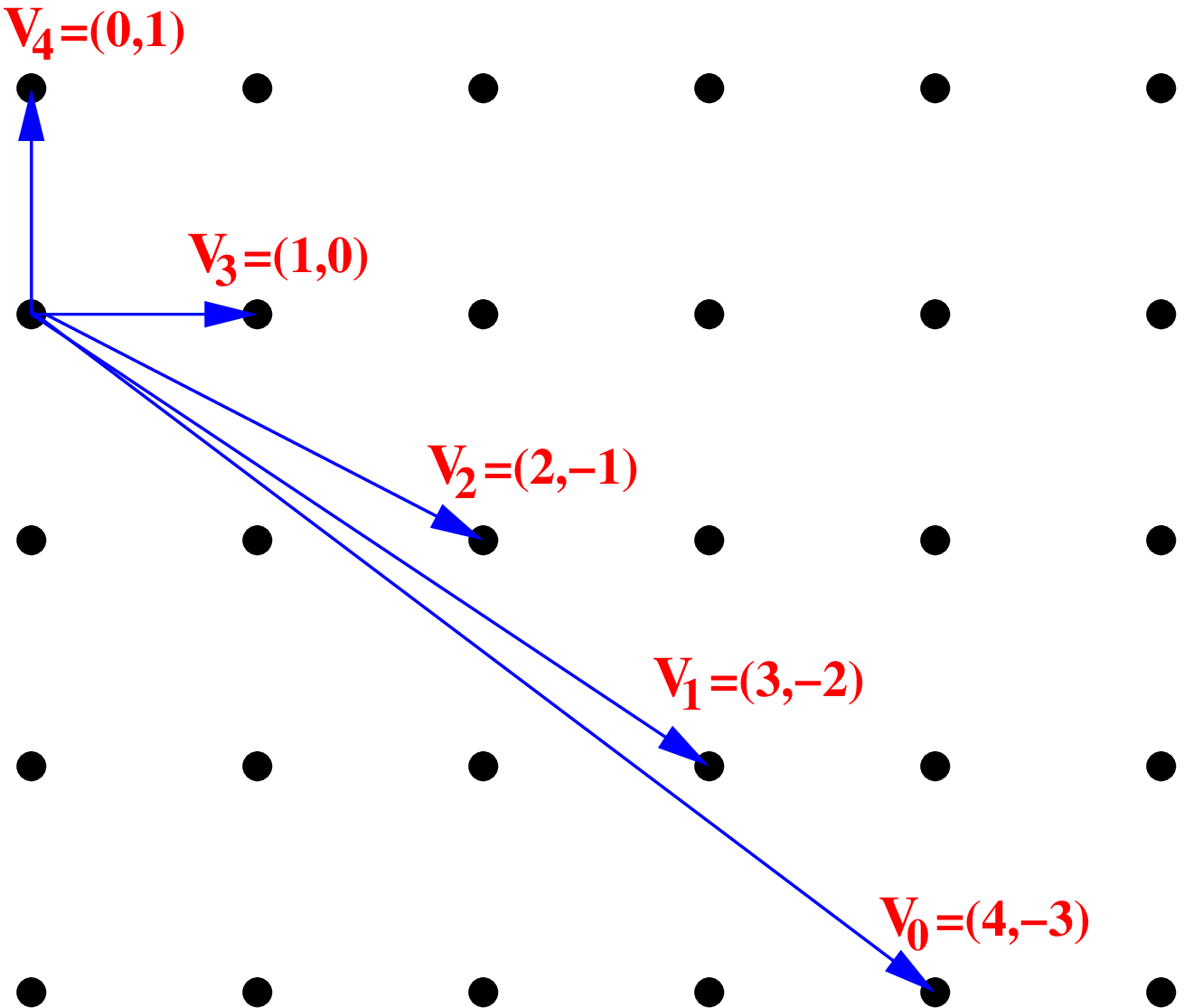}}  
        \nobreak\bigskip  
    {\raggedright\it \vbox{  
{\bf Figure 4.}  
{\it   
Toric fan for the resolution of the $\IC^2/\IZ_{n(-1)}$ 
singularity for $n=4$. 
} }}}}  
    \bigskip}           
 
Some examples which will be useful later include 
 
\item{1.} $(n,p)=(n,1)$. The continued fraction $n/1=[n]$ tells us 
that the resolved geometry has one $\IP^1$ with self-intersection  
number $-n$. 
\item{2.} $(n,p)=(n,3)$. There are two distinct cases 
depending on whether $n$ is $1$ or $2$ $\mod~ 3$. For $(3m+1,3)$ we 
have the continued fraction expansion  
\eqn\contza{{3m+1 \over 3} = [m+1,2,2]} 
while for $(3m+2,3)$ we have 
\eqn\contzb{ {3m+2 \over 3}=[m+1,3]} 
There are three or two exceptional divisors in total,  and of these  one  
has  self-intersection $-(m+1)$. 
 
\item{3.} $(n,p)=(n,-3)$. Again there are two cases, $n=3m+1$, $n=3m+2$: 
\eqn\contxa{{3m+1 \over 3m-2}= [2^{m-1},4] } 
while 
\eqn\contxb{{3m+2 \over 3m-1} = [2^{m-1},3,2]} 
There are now  $m-1$ or $m$  exceptional divisors with self-intersection $-2$ and 
one  additional  divisor with self-intersection $-4$ or 
$-3$ respectively. Note that the resolved geometry depends  
dramatically on the sign of $p$, or equivalently 
on the choice of complex structure.  
 
\noindent 
Given a fan corresponding to a non-singular variety, it is  
possible to add additional internal vectors which preserve the 
above conditions for a non-singular fan. This corresponds to blowing 
up  a non-singular point, or equivalently to adding a $\IP^1$ with self-intersection 
number $-1$. For example, the cone generated by $v_0=e_1$ and $v_1=e_2$ 
corresponds to $\IC^2$. We can add an interior vector $v_b=e_1+e_2$, 
the resulting fan describes the blow up of $\IC^2$ at a point and the 
$\IP^1$ corresponding to $v_b$ has self-intersection number $-1$ 
since $v_b=v_0+v_1$. Conversely, we can blow down such a curve and 
still find a non-singular space, or in the toric description, an interior 
vector $v_k$ which satisfies $v_{k}=v_{k-1}+v_{k+1}$ can always 
be removed from the fan while preserving the conditions for a  
non-singular variety. 
 
\subsubsec{~CFT and resolved Hirzebruch-Jung singularities} 
 
The chiral ring of an $N=2$ conformal field theory can be 
specified in terms of generators and relations.
%
Here we will summarize an empirical 
relation we have discovered
 between  the generators and relations of the ring of chiral 
twist operators \znptwist\ of the 
$\IC^2/\IZ_{n(p)}$ orbifold  
and the combinatorial data of the toric variety giving 
a minimal resolution of singularities.  
This empirical relation has been checked in a  
wide variety (including several infinite classes) 
of examples; in particular, it holds for the 
particular examples studied below. 
A more detailed exposition will appear elsewhere \hkmmtwo. 
 
The chiral ring of the orbifold is finite dimensional; we choose  
a set of generators $W_0=Y^{\sst(2)}$, 
$W_{r+1}=Y^{\sst(1)}$ with $Y^{\sst(1)}, Y^{\sst(2)}$ defined  
as in \genchir\ for each $\IC$ component of $\IC^2$, and 
$W_i$, $i=1,...,r$  constructed as products of bosonic 
and fermionic twist fields as in \znptwist. The $\{W_i\}$ are 
ordered by the twist sectors. Explicit descriptions of the $W_i$  
in some classes of examples are given in subsection 4.3 and in  
Appendix C. 
 
There is a one-to-one correspondence  
between generators of the chiral ring  
and vectors in the toric fan which describes the minimal 
resolution of the corresponding Hirzebruch-Jung singularity.  
Furthermore, the relations between the vectors  
of the toric fan \arelns\ are reflected 
in the chiral ring by relations between the corresponding generators: 
\eqn\ringrels{ 
  W_i^{a_i}=W_{i-1}W_{i+1}\quad,\qquad i=1,\ldots,r \ . 
} 
In general, these provide a subset of the full set of relations 
satisfied by the orbifold chiral ring generators. 
 
Thus we are led to the following general relation between 
the free field orbifold CFT and the corresponding resolved 
Hirzebruch-Jung singularity. The orbifold CFT corresponds to a minimal 
resolution of the singularity. There are $r$ blowing up 
parameters turned on, but as in the ALE case \AspinwallEV, 
they correspond to $B$-fields and not geometric resolutions. 
Unlike the ALE case, the operators $W_i$ are not in general 
marginal; thus the above non-trivial $B$-fields give rise to 
isolated fixed points and not lines of CFT's.  
 
Non-minimal resolutions of the singularity correspond to blowing up 
curves with self-intersection $(-1)$. A $(-1)$ curve corresponds to  
an interior vector $v_i=v_{i-1}+v_{i+1}$. The corresponding element 
of the chiral ring can be written as the product of adjacent elements,  
\ie\ it is not an independent generator. Blowing up  a $(-1)$ curve  
corresponds to perturbing by this dependent element of the chiral ring.  
More generally, perturbing the Lagrangian by
$\lambda_i\int d^2\theta\, X_i$ 
corresponds to blowing up various curves in the orbifold.
The coupling $\lambda_i$ is related to the complexified
Kahler class of this blow up.
This connection will prove useful later when we discuss RG flows.


\subsec{$\IC^2/\IZ_{n(-1)}\,:$ 
The duality between orbifolds and fivebranes}  
  
The spacetime supersymmetric orbifold theory  
$\IC^2/\IZ_{n(-1)}$ is  
T-dual to the theory of fivebranes on a circle,  
in an appropriate limit \refs{\OoguriWJ, \KutasovTE}.  
Consider type II string theory on $\IR^{8,1}\times S^1$,  
with $n$ NS5-branes symmetrically arranged on the circle,  
which we take to have circumference $R$, and parametrize by  
$x^9$; and let $x^{6,7,8}$ parametrize the $\IR^3$ transverse  
to the fivebranes (see figure 5). The claim is that in the 
limit  
\eqn\decoup{  
  g_s\rightarrow 0\quad,\qquad  
  R/l_s\rightarrow 0\quad,\qquad  
	{\rm with} \quad {R\over l_sg_s } \quad {\rm fixed}  
}  
type IIB string theory in the fivebrane background is  
equivalent to type IIA string theory on the orbifold  
$\IC^2/\IZ_{n(-1)}$ (and vice versa).

\bigskip  
{\vbox{{\epsfxsize=4in  
        \nobreak  
    \centerline{\epsfbox{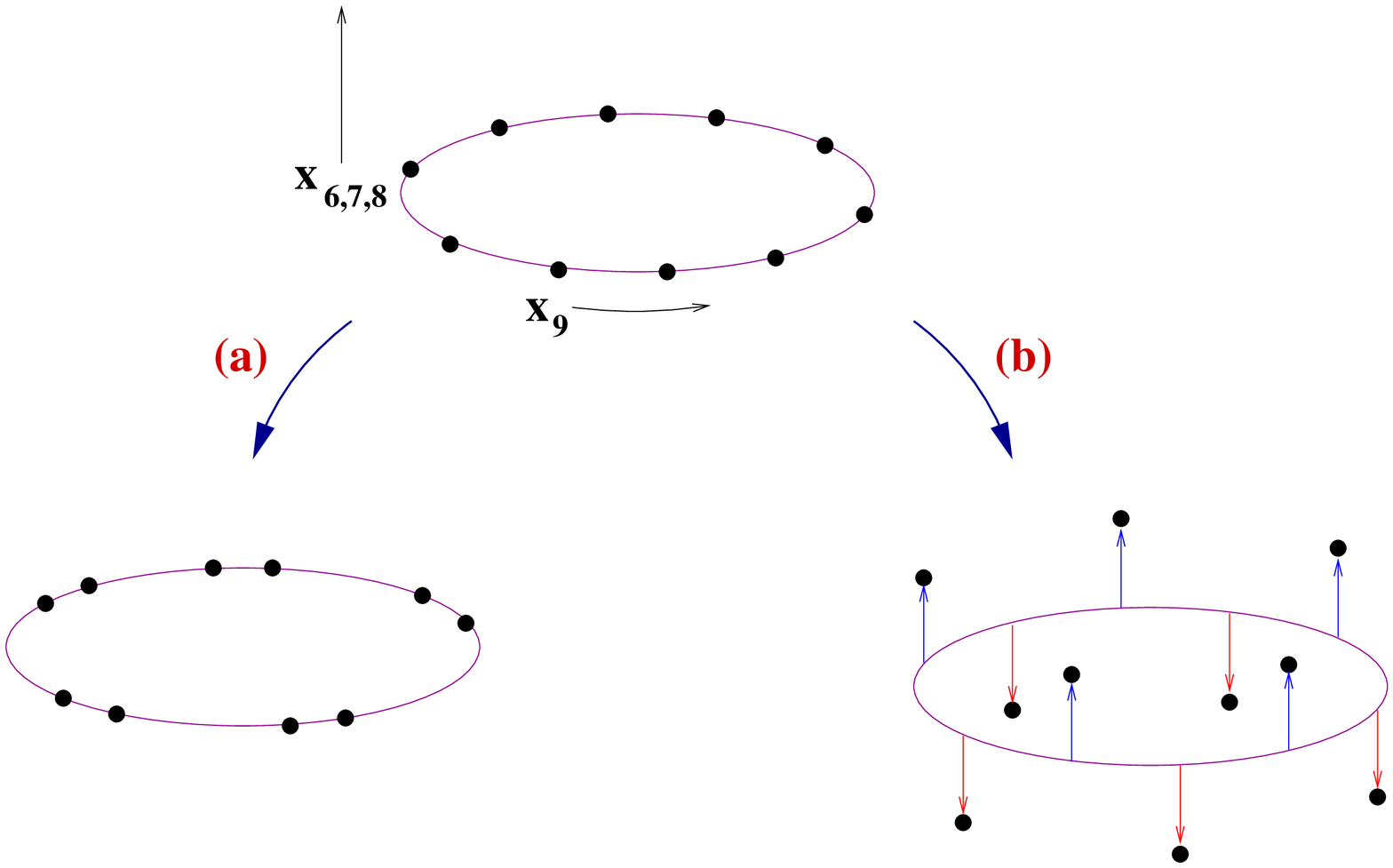}}  
        \nobreak\bigskip  
    {\raggedright\it \vbox{  
{\bf Figure 5.}  
{\it   
Two perturbations of a $\IZ_{2l}$ symmetric arrangement  
of type IIB fivebranes on a circle, dual to type IIA  
string theory on $\IC^2/\IZ_n$, with $n=2l$:  
(a) moving the fivebranes on $S^1$ is related to   
changing NS $B$-field fluxes through vanishing cycles  
on the IIA side;  
(b) moving them in $\IR^3$ is dual to turning on the triplets of  
geometrical blow up modes of the vanishing cycles   
on the IIA side.  
} }}}}  
    \bigskip}           
  
The two descriptions are related by T-duality applied to  
the circle parametrized by $x^9$. The string couplings  
are related by the standard T-duality formula  
\eqn\couprel{  
  g_s^{\sst A}=g_s^{\sst B}l_s/R\ .  
}  
Note that the limit \decoup\ is such that $g_s^{\sst A}$ is  
held fixed as $g_s^{\sst B}\to 0$. 
The orbifold has $n-1$ hypermultiplets of moduli  
coming from twisted sectors;  
the four real parameters in each hypermultiplet consist of   
the NS $B$-field flux through one of the $n-1$ vanishing  
cycles of the orbifold ALE space, together with a triplet  
of modes that blow up that cycle.  
The $B$-flux is a periodic coordinate, while the   
blow up modes parametrize $\IR^3$.  
These map on the fivebrane side into the relative locations  
of the fivebranes on the $S^1$ and $\IR^3$, respectively.  
The standard $\IC^2/\IZ_{n(-1)}$ orbifold CFT 
corresponds to the point in moduli space where the  
fivebranes are coincident in $\IR^3$ and symmetrically  
arranged on the $S^1$ (as in the top of figure 5).   
The $\IZ_n$ symmetry that cyclically permutes the fivebranes  
is the $\IZ_n$ quantum symmetry of the orbifold that   
multiplies the $j^{\rm th}$ twisted sector by $\omega^j$. 
  
There are two important classes of limits in the moduli space,  
which are easy to see in the fivebrane description (see figure 5):  
\item{(a)} Separating fivebranes along $\IR^3$ leads,  
at infinite distance, to a direct sum of decoupled theories.  
For example, if $n=2l$, we can separate the branes into two  
groups while preserving a $\IZ_l$ symmetry (figure 5b), leading  
to two decoupled $\IC^2/\IZ_l$ theories in the limit.  
\item{(b)} Bringing groups of fivebranes together   
so that they coincide at point(s) in $\IR^3\times S^1$ (figure 5a)  
leads to enhanced gauge symmetry of the fivebrane gauge  
theory.  This limit is a finite distance in the moduli  
space from the orbifold point, 
and results in a singular CFT.  
  
\noindent  
Near-coincident NS5-branes generate a target space for  
perturbative worldsheet string theory which develops  
a throat along which the string coupling grows;  
the throat becomes infinitely long, and the coupling at its end  
diverges, in the limit where fivebranes coincide  
\CallanDJ.  On the IIA side, this singularity of  
the worldsheet CFT can be understood from considerations  
of linear sigma models.  
 
One can also match the structure of D-branes on the two sides.  
The limit \decoup\ keeps fixed the mass in string units  
of D1-branes stretching between the NS5-branes on  
the IIB side; their mass scales as   
\eqn\mdone{  
  l_sm_{\sst W}=\frac{R}{n\;l_sg_s^{\sst B}}  
}  
at the point in moduli space related to the orbifold.  
D1-branes of fractional winding are pinned to the NS5-branes  
they begin and end on, while D1-branes of integer winding  
are free to move in the $\IR^3$ transverse to the NS5-branes.  
This is exactly the same structure obtained in IIA string  
theory on $\IC^2/\IZ_{n(-1)}$ \DouglasSW. There, fractional  
D0-branes of the orbifold are the W-bosons of a spontaneously  
broken 5+1 dimensional gauge symmetry localized on the orbifold  
singularity; their mass is 
\eqn\mfrac{  
  l_sm_{\sst W}=\frac1{n\;g_s^{\sst A}}\ .  
}  
These excitations are D2-branes wrapping  
the vanishing cycles of the ALE space, and carrying  
a fractional unit $1/n$ of D0-brane charge.  
Combining one fractional brane of each type makes a  
``regular'' D0-brane that can be moved off the orbifold  
point into the ambient ALE space.  
  
The $U(1)^n$ gauge fields and bifundamental matter content on 
a regular D-brane are neatly summarized in a $\IZ_{n}$ quiver  
diagram \DouglasSW. In the IIB picture, the same gauge 
theory describes the dynamics of a D1-brane wrapped around 
the circle, which intersects all $n$ NS fivebranes.  
 
As mentioned above, 
one nice feature of the fivebrane description is that it provides 
a simple geometric interpretation of the effect of closed string 
moduli. This is useful for analyzing the effects on D-branes 
of turning on various moduli. In particular, 
the relative gauge couplings of the D-brane quiver gauge theory,   
which on the IIA side are controlled by the NS B-field flux  
through the ALE vanishing cycles, correspond in the IIB description,  
to the separations of the branes on the $x^9$ circle.  
The triplets of modes that blow up the vanishing cycles to a  
smooth ALE space on the IIA side and correspond to 
Fayet-Iliopoulos D-terms on the fractional D-branes,  
are mapped in the IIB description to the relative positions  
of the branes in the $\IR^3$ parametrized by $x^{6,7,8}$.  
  
Fractionally wound branes become massless if fivebranes coincide  
(IIB), or equivalently (IIA) when the B-flux through vanishing cycles  
of the ALE space is turned off \AspinwallEV;   
the D-brane gauge dynamics then becomes strongly coupled.  
This is the open string reflection of the singularity  
of the CFT noted above. When fivebranes are infinitely 
separated in $\IR^3$, the energy of the corresponding 
fractional branes goes to infinity. 
  
The asymptotic density of localized states for the supersymmetric  
$\IC^2/\IZ_{n(-1)}$ orbifold is one half of \gnone, due to the  
GSO projection (see the discussion surrounding equations  
\diaggso, \chigso). It is interesting that $\gcl$ is not linear in  
$n$, $g_{\rm cl}\sim n^3$ for large $n$.  
This means that it is not linear in the number of fivebranes  
in the dual description. Since $g_{\rm cl}$ is constant 
under marginal deformations, it scales like $n^3$ even for  
fivebranes separated by a large but finite distance. Hence,  
it cannot be thought of as the energy of the branes, which  
would be additive for well-separated branes.

\def\th{{\rm th}}  
 
\subsec{Flows} 
 
In the previous two subsections we described some tools for studying 
the $\IC^2/\IZ_{n(p)}$ orbifold CFT's. In this subsection we will use 
these tools for analyzing the RG flows in these vacua. We  
will discuss a few special cases, and abstract from them some patterns  
which we believe are more general. A more complete analysis of the  
flows will be postponed to another publication \hkmmtwo. 
 
\subsubsec{~Example 1: $\IC^2/\IZ_{n(1)}$.} 
 
\noindent 
Our first example is the orbifold $\IC^2/\IZ_{n(1)}$. As discussed 
in section 4.1, the continued fraction expansion is in this case  
trivial, $n/1=[n]$; the chiral ring has a single  
generator $X=X_1$ of R-charge $2/n$, subject to the relation 
\eqn\znonerel{ 
  X^n=Y^{\sst(1)}Y^{\sst(2)}\ , 
} 
which is a special case of \ringrels.  
The operator $X^j$ has R-charge $2j/n$ and dimension $j/n$;  
thus, the operators with $j<n/2$, which are relevant, and  
$X^{n\over2}$ (for even $n$) which is marginal, can be added  
to the Lagrangian as perturbations of the superpotential 
\eqn\delw{ 
  \delta\LL=\lambda_j\int d^2\theta X^j~+~{\rm c.c.}\ . 
} 
The $\{\lambda_j\}$ have R-charge $1-\frac{2j}n$ and 
quantum $\IZ_n$ charge $-\frac{j}n$. 
Consequently, the chiral ring relation \znonerel\ 
cannot be modified at first order in the $\{\lambda_j\}$; 
rather the leading modification is 
\eqn\secord{ 
  X^n+\sum_{j+k<n} c_{jk}\lambda_j\lambda_k X^{j+k}=Y^{\sst(1)}Y^{\sst(2)}\ . 
}  
The flows are rather similar to the $\IC/\IZ_n$ 
examples studied in section 3. At large $\lambda$ one flows  
to the chiral ring $X^m=Y^{\sst(1)}Y^{\sst(2)}$ for  
some $m<n$. One easily checks that $\gcl$, given in 
equation \gnone, decreases along RG flows. 
 
Along the marginal line parametrized by $\lambda_{n/2}$, 
the ring relation does not change: 
\eqn\marring{ 
  X^n(1+c\lambda_{n/2}^2+...)=Y^{\sst(1)}Y^{\sst(2)}\ . 
} 
This is consistent with the expectation that $\gcl$  
is unchanged along this marginal line. 
 
The above discussion was in the context of type 0 string theory. 
The chiral GSO projection \ctwogso\ eliminates the entire  
chiral ring generated by $X$. Therefore, these flows are  
impossible in type II string theory on $\IC^2/\IZ_{n(1)}$  
(which in fact is spacetime supersymmetric).  
 
This example generalizes straightforwardly to orbifolds of 
type $\IC^m/\IZ_n$ with the group action $Z_i \sim \omega Z_i$
and $m>2$. The $g_{cl}$-conjecture in this case is verified in  
appendix A. 


\subsubsec{~Example 2: $\IC^2/\IZ_{n(-1)}$.} 
 
\noindent 
The next example we would like to analyze is $\IC^2/\IZ_{n(-1)}$, 
the spacetime supersymmetric orbifold. The continued fraction 
is $n/(n-1)=[2^{n-1}]$ (see section 4.1); thus all $n-1$ chiral  
operators $X_j$ \znptwist\ are in this case generators. 
The continued fraction expansion suggests \ringrels\ that   
$X_j^2=X_{j-1}X_{j+1}$, but in fact  
both the r.h.s. and the l.h.s. are separately zero.  
More generally, one has $X_i X_j=0$ for all $i,j=1,2,\cdots, n-1$. 
Note also that all operators $X_j$ survive the chiral GSO projection 
\ctwogso; thus the analysis of flows is the same in type  
0 and type II. 
 
The supersymmetric orbifold of course does not contain any tachyons;  
the chiral operators $X_j$ correspond to massless states. As discussed 
above, one can still ask what happens when we perturb the Lagrangian 
by a marginal operator $\lambda_j\int d^2\theta X_j$, and send the  
modulus $\lambda_j\to\infty$.  
 
Geometrically, turning on $\lambda_j$ and sending it to infinity 
corresponds to blowing up the appropriate $\IP^1$ in the ALE geometry 
to infinite size. In this limit, $\lambda_j$ can be thought of as 
a Lagrange multiplier imposing the constraint $X_j=0$; the 
corresponding infinite size $\IP^1$ disappears from the singular 
geometry. The generators $X_i$ with $i<j$ decouple from those with 
$i>j$; the corresponding $\IP^1$'s do not intersect (see figure 3). 
 
All this can be summarized as the following action on the  
Hirzebruch-Jung continued 
fraction: 
 
\eqn\splitsusy{ 
  [2^{n-1}]\to [2^{j-1}]\oplus[2^{n-j-1}] 
} 
where we erased the $j^\th$ entry in the continued  
fraction and took into account the decoupling of  
the left and right parts of the fraction. 
Eq. \splitsusy\ describes a process in which a $\IZ_n$  
singularity is split 
into decoupled $\IZ_j$ and $\IZ_{n-j}$ singularities: 
\eqn\susyspl{ 
\IC^2/\IZ_{n(-1)}\to \IC^2/\IZ_{j(-1)}\oplus \IC^2/\IZ_{n-j(-1)}\ .} 
This process is easily understood in terms of fivebranes (see section 
4.2). It corresponds to a deformation that takes $n$ fivebranes  
arranged symmetrically on a circle (as in the top figure in figure 5) 
and moves $j$ of them in the $(6,7,8)$ directions such that the 
final configuration has two groups of $j$ and $n-j$ fivebranes,  
infinitely separated in $\IR^3$. It is easy to check that, as  
in the previous example, $g_{\rm cl}$ \gnone\ decreases in the process. 

The discussion above can be generalized to any flow
corresponding to a marginal or relevant generator of the
chiral ring in any $\IC^2/\IZ_{n(p)}$ CFT. Adding the generator
$W_j$ with a large coefficient\foot{Note that unlike
the situation for marginal deformations, for relevant
deformations one does not need to send the coupling to
infinity by hand; it grows naturally along the RG flow.}
has the following effect on the continued fraction:
\eqn\splitchi{
  [a_1,\ldots,a_r]\to [a_1,\ldots,a_{j-1}]\oplus[a_{j+1},\ldots,a_r]\ .
}
The flow thus splits the target space into a direct sum of
the corresponding Hirzebruch-Jung singularities.
  

\subsubsec{~Example 3: $\IC^2/\IZ_{2\l(-3)}$.} 
 
\noindent 
Our next example is $\IC^2/\IZ_{2\l(-3)}$. As mentioned in section 4.1 
(equations \contxa, \contxb), the continued fraction is slightly different 
for the cases $n=3m+1$ and $n=3m+2$. 
\eqn\caseii{\eqalign{ 
  [\,\underbrace{2,\ldots,2}_{m-1}\,,4] &\quad,\qquad n=2\l=3m+1\cr 
  [\,\underbrace{2,\ldots,2}_{m-1}\,,3,2] &\quad,\qquad n=2\l=3m+2\ . 
}} 
For $n=3m+1$, the $m$ generators of the chiral ring are  
$W_j=X_j$ with $j=1,2,\cdots, m$, with R-charges \rchtwist\ 
$R_j=1-{2j\over n}$. For $n=3m+2$ there are $m+1$ generators,  
the first $m$ of which are as in the previous case, while 
$W_{m+1}=X_{2m+1}$, whose R-charge is $R_{2m+1}=2-{2\over n}(2m+1)$ 
(see Appendix C for further details). 
All generators are relevant operators. Chiral GSO takes  
$W_j\to -W_j$ for $j=1,..., m$ and for $n=3m+2$, 
$W_{m+1}\to W_{m+1}$. Thus, the generators $W_1,...,W_m$ 
survive the GSO projection, while $W_{m+1}$ is projected out. 
 
One can now study various relevant and marginal perturbations of 
the CFT. Consider first deformations by the generators.  
For $n=3m+1$, condensing $W_j$ with $1\le j\le m-1$ leads,  
as in the previous example, to the flow 
\eqn\flgenone{[2^{m-1},4]\to [2^{j-1}]\oplus [2^{m-j-1},4]\ ,} 
or equivalently: 
\eqn\flgentwo{\IC^2/\IZ_{3m+1(-3)}\to \IC^2/\IZ_{j(-1)}\oplus 
\IC^2/\IZ_{3(m-j)+1(-3)}\ .} 
This is an example where a non-supersymmetric orbifold produces 
 a lower rank non-supersymmetric orbifold as well as 
a supersymmetric one. By plugging in the explicit formulae for 
$g_{\rm cl}$ \gnone, \gnthree\ one finds that, as expected: 
\eqn\gtheor{ 
g_{\rm cl}(3m+1, -3)>g_{\rm cl}(j,-1)+g_{\rm cl}(3(m-j)+1, -3)\ .} 
Perturbing by $W_m$ gives rise to the flow 
\eqn\wmflow{\eqalign{[2^{m-1},4]&\to[2^{m-1}]\cr 
\IC^2/\IZ_{3m+1(-3)}&\to \IC^2/\IZ_{m(-1)}\ . 
}} 
Again, one can check that $g_{\rm cl}$ decreases along the flow. 
A similar set of flows is obtained when one perturbs by generators 
of the chiral ring in the case $n=3m+2$. Note that all the 
flows described above in the type 0 context exist in the type II 
theory as well, since the generators by which we perturbed 
survive the GSO projection. 
 
A new element in this case is that one can perturb by products 
of generators. For example, consider the chiral operator 
\eqn\newgen{ 
  V_j=W_jW_{j+1}\  
} 
with $(n-1)/4\le j\le m-1$. One can show that $V_j$ is a non-vanishing 
relevant (or marginal) operator. Adding it to the worldsheet 
Lagrangian corresponds in the Hirzebruch-Jung geometry to blowing up the 
point of intersection of the $j^\th$ and $(j+1)^{\rm st}$  
$\IP^1$'s in the minimal resolution of the singularity.  
In general, when one blows up such a $\IP^1$ in a resolved 
Hirzebruch-Jung surface, the effect on the continued fraction is 
the following (see \fulton\ p. 44), 
\eqn\blowupch{[a_1,\cdots, a_r]\to 
  [a_1,\ldots,(a_j+1)\,,1,\,(a_{j+1}+1),\ldots,a_r]\ . 
} 
In the CFT this corresponds to perturbing by the operator 
$V_j$ \newgen\ and adding it to the set of ring generators, 
together with the trivial ring relation \newgen.  
Note that the continued fractions \contfrac\ corresponding 
to the l.h.s. and r.h.s. of \blowupch\ are equal.  
 
Sending the coefficient of $V_j$ in the action to infinity 
can now be treated in the same way as before. The continued 
fraction is split into two disconnected components, 
\eqn\splitchii{ 
  [a_1,\ldots,(a_j+1)\,,1,\,(a_{j+1}+1),\ldots,a_r] 
	\to [a_1,\ldots,(a_j+1)]\oplus[(a_{j+1}+1),\ldots,a_r] 
} 
and the CFT correspondingly splits into a direct sum of the 
appropriate orbifolds.  
 
In the examples \caseii\ considered here, perturbing by 
\newgen\ thus has the following effect: 
\eqn\caseiidaughters{\eqalign{[\,\underbrace{2,\ldots,2}_{m-1}\,,4] 
  \to[\,\underbrace{2,\ldots,2}_{j-1}\,,3]&\oplus 
	[3,\,\underbrace{2,\ldots,2}_{m-2-j}\,,4]\cr 
  [\underbrace{2,\ldots,2}_{m-1}\,,3,2]\to 
[\,\underbrace{2,\ldots,2}_{j-1}\,,3]&\oplus 
	[3,\,\underbrace{2,\ldots,2}_{m-2-j}\,,3,2]\ . 
}} 
 
\subsubsec{~Example 4: $\IC^2/\IZ_{2\l(3)}$.} 
 
\noindent 
The last set of examples that we will consider here is the 
case $\IC^2/\IZ_{2\l(3)}$, which was studied in \AdamsSV.  
These authors found that under a large marginal deformation, 
this model exhibits the flow 
\eqn\twolthree{\IC^2/\IZ_{2\l(3)}\to \IC^2/\IZ_{\l(1)}\oplus 
\IC^2/\IZ_{\l(-3)}\ .} 
If true, this provides a counterexample to the 
$g_{\rm cl}$-conjecture. For example, for large $\l$ one 
has 
\eqn\gcltwol{ 
\eqalign{ 
g_{\rm cl}(2\l,3)=&{1\over 720} {(2\l)^3\over 9}\cr 
g_{\rm cl}(\l,1)+ g_{\rm cl}(\l,-3)=&{1\over 720}\left(\l^3+{\l^3\over9} 
\right)\cr 
}} 
and $g_{\rm cl}(2\l,3)<g_{\rm cl}(\l,1)+ g_{\rm cl}(\l,-3)$. 
%
We will next examine this model in some detail. We will argue 
that the flow \twolthree\ is not possible in CFT if we interpret the 
right-hand side of \twolthree\ as a pair of free field orbifolds. 
 
Let $n=2\l=3m+2$ and $p=3$; the continued fraction is particularly 
simple in this case, 
\eqn\znthreecf{ 
  \frac n3=[m+1,3]\ . 
} 
The generators of the chiral ring are $W_1=X_1$ 
and $W_2=X_{m+1}$. They have R-charges \rchtwist\ 
$R_1=4/n=2/\l$ and $R_2=(m+2)/n$, and satisfy the relations  
$W_1^{m+1}=W_2^3=0$. The chiral GSO projection takes $W_1\to W_1$, 
$W_2\to -W_2$. The chiral ring consists of three bands: 
\eqn\threebands{\eqalign{ 
  {\rm band~1}~&:\qquad W_1,W_1^2,\ldots,W_1^m\cr 
  {\rm band~2}~&:\qquad W_2,W_1W_2,\ldots,W_1^mW_2\cr 
  {\rm band~3}~&:\qquad W_2^2,W_1W_2^2,\ldots,W_1^{m-1}W_2^2\ . 
}} 
Only band 2 survives the chiral GSO projection. 
As a side remark we note that following our previous 
analysis it is easy to understand the perturbations 
\eqn\simplepert{\delta\CL=\int d^2\theta\left(\lambda_1W_1 
+\lambda_2 W_2+\lambda_3 W_1 W_2\right)\ .} 
Turning on $\lambda_1$ takes 
\eqn\oneflow{[m+1,3]\to [3]\ ;} 
Perturbing by $\lambda_2$ takes 
\eqn\twoflow{[m+1,3]\to [m+1]\ ;} 
$\lambda_3$ takes  
\eqn\threeflow{[m+1,3]\to [m+2]\oplus[4]\ .} 
Due to the transformation properties of $W_j$ under the chiral 
GSO projection, the flow \oneflow\ can only be realized in the  
type 0 theory, while the other two flows exist in the type II  
theory as well.  
 
The marginal perturbation studied in \AdamsSV\ corresponds to  
\eqn\margpert{\delta\CL=\lambda\int d^2\theta \;W_2W_1^{m\over2}} 
It does not seem to fit into the class of perturbations 
discussed above, but it can be brought to that form by a 
sequence of blow ups of $\IP^1$'s with self intersection $-1$. 
Such blow ups are described by equation \blowupch.   
The first blow up gives 
\eqn\oneblow{\eqalign{ 
  [m+1,3]&\to[m+2,1,4]\cr 
  \st [W_1,W_2]&\to\st [W_1,W_1\!W_2,W_2]\ , 
}} 
where on the second line we have written the generators  
in the chiral ring corresponding to the different entries 
in the continued fraction, including the $-1$ curve that  
is being blown up (here ${\st W_i}=W_i$). Successive blow 
ups yield 
\eqn\moreblow{\eqalign{ 
  [m+2,1,4]\to[m+3,1,2,4]&\to\cdots 
	\to[m+j,1,\,\underbrace{2,\ldots,2}_{j-2}\,,4]\cr 
  {\st [W_1,W_1\!W_2,W_2]}\to\st [W_1,W_1^2\!W_2,W_1\!W_2,W_2]&\to\cdots 
	\to\st [W_1,W_1^{j-1}\!W_2,\, 
	\underbrace{\st W_1^{j-2}\!W_2,\ldots,W_1\!W_2}_{j-2}\,,W_2]\ . 
}} 
Geometrically, after blowing up the $j-1$ $\IP^1$'s in \moreblow\ 
one can ask what happens when the radius of the $\IP^1$ corresponding 
to $V=W_1^{j-1}\!W_2$ (say) is sent to infinity. 
Following our previous analysis, one would say that  
the geometry splits into the pair of Hirzebruch-Jung singularities 
whose continued fractions are 
\eqn\infinitivesplit{ 
  [m+1,3]\to [m+j,1,\underbrace{2,\ldots,2}_{j-2}\,,4]\to 
	[m+j] \oplus [\,\underbrace{2,\ldots,2}_{j-2}\,,4]\ , 
} 
which we recognize as the pair 
$\IC^2/\IZ_{m+j(1)}\oplus\IC^2/\IZ_{3j-2(-3)}$. 
In particular, for $j={m\over2}+1$  one recovers the  
result of \AdamsSV, \twolthree. 
 
In the CFT, the blow up described above is supposed
to correspond to perturbing the $\IC^2/\IZ_{2\l(3)}$ CFT by  
\eqn\margnew{\delta\CL=\lambda_{j-1}\int d^2\theta \;W_2W_1^{j-1}} 
with $j=3,4,\cdots, {m\over2}+1$. For $j\le {m\over2}$ the 
perturbation is relevant, while for $j={m\over 2}+1$ it  
reduces to \margpert\ and is marginal.  
For large enough $j$, the process \infinitivesplit\
violates the $\gcl$-conjecture.
We would next like to argue that it does not occur in CFT.
 
Consider first the case where $\lambda_{j-1}$ is a relevant  
perturbation. In order to analyze the perturbation \margnew\ 
one can proceed as indicated in \infinitivesplit. One first 
blows up by a small amount $j-2$ cycles, and by a much larger 
amount the cycle corresponding to \margnew, and then flows to the 
IR on the worldsheet (thereby sending $\lambda_{j-1}$ and the radius 
of the cycle to infinity). In order to focus on the perturbation  
\margnew\ one would like to send the radii of the other $j-2$ cycles  
to zero.  
 
The $j-2$ blowing up parameters correspond to coefficients in 
the Lagrangian of $W_1W_2, W_1^2W_2,\cdots, W_1^{j-2}W_2$. These 
operators are relevant, and thus their coefficients in the Lagrangian 
can really be thought of as a series of (energy) scales  
$\mu_1, \cdots, \mu_{j-2}$. The coupling $\lambda_{j-1}$  
also gives rise to a scale, $\mu_{j-1}$, and one can  
study the situation in which $\mu_{j-1}\gg \mu_k$,  
$k=1,2,\cdots, j-2$. The original problem, of analyzing \margnew\ 
corresponds to sending $\mu_k\to 0$.  
 
{}From the point of view of the geometric analysis \infinitivesplit,  
the limit $\mu_k\to 0$ is singular. For the transition 
\eqn\splittlt{\IC^2/\IZ_{2l(3)}\to 
\IC^2/\IZ_{m+j(1)}\oplus\IC^2/\IZ_{3j-2(-3)} 
} 
to occur, it was important that the parameters $\mu_k$ are kept 
finite. But if the $\mu_k$ are finite,   
as the system evolves to the IR it will eventually probe the 
region $E\ll \mu_k$, where the geometry is no longer described 
by the r.h.s.'s of equations \infinitivesplit, \splittlt. 
If we try to take the scales  
$\mu_k$ to zero, the geometric analysis seems to 
suggest that we find a singular CFT. 
Therefore, it is not clear 
to us that there is a limit in which the CFT realizes the flow 
\splittlt. 
 
We next turn to the case where $\lambda_{j-1}$ is a marginal 
operator, \margpert. The discussion here is similar to the above, 
except $\lambda$ is now a dimensionless parameter. In order to 
arrive at the flow \twolthree\ by using \infinitivesplit\ 
one again has to blow up $m/2$ $\IP^1$'s 
which correspond in the CFT to the relevant operators 
$W_2,W_1W_2,\cdots, W_1^{{m\over2}-1}W_2$. This introduces 
$m/2$ scales $\mu_k$, $k=1,2,\cdots, m/2$. If one attempts to 
send the modulus $\lambda\to\infty$ while sending the scales  
$\mu_k\to0$, one again seems to find a singular CFT, while 
if one keeps the scales $\mu_k$ finite, the RG flow eventually 
makes them very large and the analysis that led to \twolthree\ 
breaks down.  
 
One could also analyze this example in the following way. 
We first consider the orbifold $\IC^2/\IZ_2$. This theory 
has a twisted marginal operator; perturbing by it gives 
an Eguchi-Hanson space, $T^* \IP^1$, with the modulus  
$\lambda$ controlling the size of the $\IP^1$.  
Choose complex coordinates $x_\pm$ on the northern and 
southern hemispheres of $\IP^1$. Cotangent vectors are 
parametrized by $p_+ dx_+ = p_- dx_-$, so the transition 
functions are 
\eqn\transf{ 
\eqalign{ 
x_+ & = 1/x_- \cr 
p_+ &  = - p_- x_-^2\ .} 
} 
The coordinates $(Z^1,Z^2)$ of $\IC^2/\IZ_2$  
are related to those of the resolved space $T^*\IP^1$ via 
\eqn\resmap{ 
\eqalign{ 
x_+ = Z_1/Z_2 \qquad & \qquad p_+ = Z_2^2\cr 
x_- = Z_2/Z_1 \qquad & \qquad p_- = - Z_1^2\ .} 
} 
We can then act with $\IZ_\l$ on the resolved space 
in the way induced by this map from the
action of $\IZ_{2\l(3)}$ on $\IC^2$. 
Since $\IC^2/\IZ_{2\l}=(\IC^2/\IZ_2)/\IZ_\l$ 
(for $\l\in 2\IZ+1$), the resulting space is the same 
as one would get by quotienting by the full $\IZ_{2\l}$, 
and then resolving. 
In the limit of large radius of the $\IP^1$ one finds 
in this way two ``daughter'' singularities 
at the north and south poles of the 
$\IP^1$ which look locally like $\IC^2/\IZ_{\l(1)}$ and  
$\IC^2/\IZ_{\l(-3)}$. Assuming that these singularities  
are described by standard free field orbifolds leads to  
\twolthree. 
 
As mentioned above, we believe that in fact the daughter 
singularities do not correspond to standard orbifold CFT's.  
Indeed, our results on the relation of the Hirzebruch-Jung  
resolution to the chiral rings of the orbifolds strongly 
suggests the following picture.  
 
The daughter conformal field theories $\CD_i$, $i=1,2$ 
are linear sigma models with target space {\it metric}  
which is the flat metric on $\IC^2/\IZ_{\ell(r_i)}$, 
with $r_1=1$, $r_2=-3$.  
However to define the theories one must also specify the  
$B$-fields. It is likely that the $B$-fields for the  
daughter conformal field theories $\CD_i$ {\it differ}  
from those of the standard free field orbifolds 
$\IC^2/\IZ_{\ell(r_i)}$. In particular, if the  
$B$-fields are such that one or both of the daughters 
corresponds to a singular CFT, the decoupling between the 
North and South poles might break down due to the appearance 
of a throat \CallanDJ. In addition, the strong coupling region 
in the throat would invalidate the CFT analysis.  
 
{}From the discussion of section 4.1 it is in fact natural 
to expect that the marginal deformation \margpert\ leads to 
a singular CFT. The reason is that as we have seen
in section 4.1.2, to reach a 
non-singular CFT one needs to blow up the $\IP^1$'s  
\infinitivesplit, but in the flow \margpert\ these 
parameters are tuned to zero.  
 
So where is the decoupled theory on the r.h.s. of \twolthree\ 
in the space of 2d field theories?  It is obtained by 
adding half a unit of $B$ flux on the $\coeff m2-1$ collapsed 
cycles at the north and south poles of $T^*\IP^1$, 
and then sending $\lambda$ to infinity
keeping the $B$ flux fixed.  The limiting 
theory cannot be reached from any finite point on the 
marginal line by an RG flow, for the reasons 
given below equation \splittlt.

\lref\WittenYC{ 
E.~Witten, 
``Phases of N = 2 theories in two dimensions,'' 
Nucl.\ Phys.\ B {\bf 403}, 159 (1993) 
hep-th/9301042. 
}

\lref\WittenBF{ 
E.~Witten, 
``Elliptic Genera And Quantum Field Theory,'' 
Commun.\ Math.\ Phys.\  {\bf 109}, 525 (1987). 
} 
 
It would be very interesting to  test this picture more thoroughly.   
Some tools which might prove useful in doing this include  the 
gauged linear sigma model of \WittenYC\ and the elliptic genus 
\WittenBF. 
 
To summarize, we believe that the geometrical transition  
\twolthree\ might not be possible in CFT, in agreement  
with the $g_{\rm cl}$-conjecture. 
 
\subsubsec{~The general structure} 
 
While most of the discussion in this subsection was done in  
special cases, it is clear that the structure generalizes 
for all $p$. We will leave a detailed description of 
the general structure to another publication \hkmmtwo\ and 
restrict here to a few brief comments.  
 
The chiral ring, which always contains $n-1$ twisted sector  
operators, splits in general into $|p|$ bands, labeled by  
$[|p|j/n]$. For large $n$ and small positive $p$, the  
structure is similar to that described above for the cases  
$p=1$, and $p=3$. There is a single generator of the ring  
in the first band of the chiral ring, with the rest of the  
operators in this band being powers of this generator. 
Similarly, in higher bands there is at most one new generator 
per band. 
 
For small negative $p$, the structure is like that described above 
for $p=-1$, and $p=-3$. All the operators in the first band are 
generators, and there are sometimes generators in higher bands  
as well. The structure of the chiral ring in general is best 
described in terms of toric geometry (section 4.1).

\newsec{Fivebranes and Liouville flows}  
  
As described in section 4, twisted sector RG flows   
in $\IC^2/\IZ_n$ orbifold models are related to the dynamics  
of NS5-branes on a small transverse circle. The physics   
of fivebranes on a large (or non-compact) transverse  
space is rather different, and thus provides additional  
examples of the phenomena explored in this paper. In  
this section we briefly discuss these examples.  
  
We start with a discussion of the supersymmetric  
system of $n$ parallel NS5-branes. We take the fivebranes  
to be localized on an $\IR^4$ labeled by $(x^6, x^7, x^8, x^9)$,  
and extended in the remaining $5+1$ directions. This system has   
no infrared instabilities (tachyons), but just as   
in section 4 we can study its behavior as a function of   
the moduli, which are the locations of the fivebranes in $\IR^4$.  
  
As a representative example, consider a system of $n$  
coincident fivebranes, and turn on moduli that correspond to  
moving $n-n'$ fivebranes to different locations in $\IR^4$,  
which are then sent to infinity. Following the discussion of   
the previous sections, one expects the number of degrees of   
freedom localized on the fivebranes to decrease in the process.   
  
Systems of coincident fivebranes are in general outside of   
the realm of applicability of our analysis. Indeed, it is   
well known \CallanDJ\ that they develop a throat with a   
linear dilaton along it, and the string coupling diverges as   
one approaches the fivebranes.  The high energy density of   
states of the system is nevertheless known, from thermodynamic   
considerations. It grows as   
\eqn\sahak{\rho(E)\sim E^\alpha\exp(\beta_HE)}  
where $\beta_H=2\pi\sqrt{n\alpha'}$, and $\alpha$ is  
a known constant \KutasovJP. Thus, the spectrum of states  
exhibits Hagedorn growth, with a Hagedorn temperature  
determined by the number of fivebranes. It should be  
emphasized that most of the states contributing to $\rho(E)$   
are expected to be non-perturbative. Also, $E$ here is the  
energy in spacetime, while in previous formulae, such as   
\mmm, it was energy on the worldsheet.  
  
What happens when the fivebranes are separated? For finite  
separations, it is clear that at energies much higher  
than the mass of the W-bosons of the broken gauge symmetry  
on the fivebranes\foot{We are discussing the type IIB  
fivebrane case, for concreteness. A similar analysis holds  
for the IIA case.}, the system is insensitive to the existence  
of the separation, and the density of states is the same  
as at the origin of the Coulomb branch \sahak. However, if  
we first send $n-n'$ fivebranes to infinity (and thus take the 
corresponding $m_W\to\infty$), the effective number of fivebranes   
decreases from $n$ to $n'$, and the corresponding  
density of states is now given by \sahak\ but with a  
smaller $\beta_H=2\pi\sqrt{n'\alpha'}$. Thus, we conclude  
that it is indeed true for fivebranes that ``flows'' in which  
some moduli are sent to infinity decrease the number of states  
of the system, as one would expect.

In order to connect to the discussion of the previous  
sections, we would like to construct a fivebrane configuration  
which can be described by weakly coupled string theory. It is  
known how to do that by going to the Coulomb branch of  
the fivebrane theory. It was shown in \refs{\OoguriWJ,  
\GiveonPX, \GiveonTQ} that the near-horizon geometry of a system   
of $n$ fivebranes symmetrically arranged on a circle of radius   
$r_0$ in $\IR^4$ is  
\eqn\fiveres{\IR^{5,1}\times   
\left({SL(2)_n\over U(1)}\times {SU(2)_n\over U(1)}\right)/\IZ_n  
}   
The radius of the circle on which the fivebranes are arranged,  
$r_0$, determines the string coupling at the tip of the cigar  
$SL(2)/U(1)$. As $r_0\to\infty$, the string coupling at the tip  
of the cigar goes to zero.   
  
The geometry \fiveres\ provides a holographically dual description  
of the dynamics of the fivebranes, also known as Little  
String Theory \refs{\AharonyUB,\GiveonPX}. Normalizable states  
in the geometry \fiveres\ correspond to states in LST;   
non-normalizable states give rise to off-shell observables.  
Actually, in the geometry \fiveres\ one finds three classes  
of vertex operators:  
\item{(1)} Normalizable states living near the tip of the cigar,  
corresponding to principal discrete series representations of  
$SL(2)$.  
\item{(2)} Delta function normalizable states living in the  
bulk of the cigar \fiveres. These correspond to the principal  
continuous series representations of $SL(2)$. They form a   
continuum above a finite  
energy gap, and are in a sense intermediate between states  
bound to the fivebranes and states that propagate in the full  
ten dimensional spacetime far from the branes.   
\item{(3)} Non-normalizable vertex operators, whose wavefunctions  
are supported in the weak coupling region far from the tip of  
the cigar. As mentioned above, these correspond to off-shell  
observables.   
  
\noindent  
Since string theory in the background \fiveres\ can be made  
arbitrarily weakly coupled by tuning the string coupling  
at the tip of the cigar, or equivalently the radius of  
the circle on which the fivebranes are placed, one can  
study this system using our techniques.  
  
To evaluate the density of localized states, one simply   
computes the torus partition sum in the background \fiveres.  
It is not difficult to show that the high energy density of states  
grows like \asdens\ with   
\eqn\ceffliouv{c_{\rm eff}=6-{6\over n}}  
This follows\foot{Note also that $c_{\rm eff}$ is additive for   
products of CFT's, and that the $\IZ_n$ orbifold in \fiveres\ cannot  
change it (it only influences the prefactor $g$).}  
from the fact that the cigar CFT has  
$c_{\rm eff}=3$ (this is analogous to the fact that Liouville  
CFT has $c_{\rm eff}=1$ for all values of the central charge  
\KutasovSV), while the $SU(2)_n/U(1)$ CFT is an $N=2$  
minimal model with $c_{\rm eff}=c=3-(6/n)$.   
  
Marginal flows that separate the fivebranes  
in the plane containing the circle which defines the original  
configuration are particularly simple to study using this   
approach. The $SU(2)/U(1)$ factor in the background \fiveres\  
can be represented by an $N=2$ Landau-Ginzburg system for  
a superfield $\chi$, with a superpotential  
\eqn\ntwomin{W=\chi^n}  
The cigar CFT $SL(2)/U(1)$ is equivalent \GiveonPX\ to $N=2$  
Liouville, a chiral superfield $\Phi$ with the superpotential  
\eqn\ntwoliouv{W=\mu e^{-{1\over Q}\Phi}}  
where $Q^2=2/n$ is the linear dilaton slope in the Liouville  
direction $\phi={\rm Re}\;\Phi$, and $\mu$ determines the string  
coupling at the tip of the cigar (and thus is related to the  
radius of the circle in the original fivebrane geometry).   
The moduli space of fivebranes is explored by deforming the  
superpotential \ntwomin, \ntwoliouv\ to  
\eqn\supgen{W=\chi^n+\mu e^{-{1\over Q}\Phi}+  
\sum_{j=2}^{n-1}\lambda_j\chi^{n-j}e^{-{Q\over2}j\Phi}}  
Note that all terms in \supgen\ are invariant under the   
$\IZ_n$ symmetry in \fiveres, which acts as   
\eqn\znsym{\chi\to e^{2\pi i\over n}\chi;\;\;  
\Phi\to\Phi-2\pi iQ;\;\;\theta\to e^{2\pi i}\theta;\;\;\bar\theta  
\to\bar\theta}  
The parameters $\lambda_j$ in \supgen\ determine the locations  
of the fivebranes. It is important to emphasize that unlike the  
formally similar situation studied in section 3 (see \eg\ equation  
\deformsup), here $\lambda_j$ are exactly marginal perturbations,  
reflecting the fact that parallel fivebranes do not exert a net  
force on each other.  
  
The main features of the moduli space can be read  
directly from \supgen,  
as in \refs{\ZamolodchikovDB,\KastorEF}.   
For finite $\{\lambda_j\}$, the large $\chi$ behavior of  
the superpotential is dominated by the $\chi^n$ term. This  
implies that the high energy density of states is governed by   
\ceffliouv\ for all finite $\{\lambda_j\}$. If some or all of  
the $\{\lambda_j\}$ are taken to infinity, the asymptotic behavior  
of the potential changes, and $c_{\rm eff}$ decreases accordingly.  
For example, if $\lambda_{n-n'}$ is sent to infinity  
$(1\le n'\leq n-2)$  
while keeping all other $\lambda_j=0$, the system decays into  
one with $n$ replaced by $n'$. This is the Landau-Ginzburg  
description of the flow in which $n-n'$ fivebranes are sent 
to infinity.   
  
A few comments about the preceding discussion are in order here:  
\item{(1)} In analyzing the behavior of \supgen, we focused on  
the dynamics of $\chi$, since in systems of $(N=2)$ matter coupled  
to $(N=2)$ Liouville, the RG flow of the matter system is exhibited  
by the full system as evolution as a function of  
$\phi(={\rm Re}\;\Phi)$ \KutasovPF.  
  
\item{(2)} The behavior of \supgen\ as a function of the moduli  
$\{\lambda_j\}$ illustrates a phenomenon we encountered in section  
4. In the discussion of $\IC^2/\IZ_n$ regular orbifolds we argued that  
the high energy density of states $(c_{\rm eff}$ and $g_{\rm cl}$)  
does not change along moduli spaces associated with localized   
perturbations, but {\it can} decrease (in that case it was only  
$g_{\rm cl}$ that changed) if we go an infinite distance in  
moduli space. The fivebrane system \supgen\ provides an explicit  
demonstration of this: for finite values of the moduli $\{\lambda_j\}$,  
the high energy density of states remains unchanged, while when  
some of the $\{\lambda_j\}$ become infinite, $c_{\rm eff}$ decreases.  
  
\item{(3)} It was seen in section 4 that for  
$n$ fivebranes arranged in two widely separated groups of  
$n_1$ and $n_2$ branes, $g_{\rm cl}$ was {\it not} approximately  
equal to $g_{\rm cl}(n_1)+g_{\rm cl}(n_2)$. While this was found   
for fivebranes in the regular orbifold limit \decoup, we see  
the same phenomenon here. In the unperturbed system \supgen\  
with $\{\lambda_j=0\}$, the density of perturbative localized   
states is independent of $\mu$, which determines the radius $r_0$ 
of the circle that the $n$ fivebranes live on.  What changes as  
$r_0\to\infty$ is the coupling between these states (which goes to  
zero). It is natural that the same happens for $\IC^2/\IZ_n$ orbifolds:  
as we tune moduli, $g_{\rm cl}$ does not change, but as the  
moduli are sent to infinity the couplings of some of the states  
to others go to zero.   
  
\item{(4)} As mentioned in the beginning of this section, the  
physics of fivebranes in a large transverse space is quite  
different from that of the regular orbifolds $\IC^2/\IZ_n$, which  
correspond to fivebranes on a small transverse circle.   
In particular, while for $\IC^2/\IZ_n$ one has 
$c_{\rm eff}=6$, for fivebranes  
it is given by \ceffliouv\ (\ie\ it is smaller). A related fact  
is that in flows between regular orbifolds, $c_{\rm eff}$ remains  
fixed, and $g_{\rm cl}$ is changing, while flows between fivebrane  
systems typically involve a change in $c_{\rm eff}$. One can ask  
why all this is not in contradiction with the  
``$g_{\rm cl}$-conjecture'', which presumably implies that both  
$c_{\rm eff}$ and $g_{\rm cl}$ are constant along marginal  
deformations -- after all, $\IC^2/\IZ_n$ and \fiveres\ are related  
by adjusting moduli. A possible answer is that to go from $\IC^2/\IZ_n$  
to \fiveres\ one has to: (a) change the value of an untwisted modulus 
(the size of the circle the fivebranes live on); (b) go an infinite  
distance in moduli space. Both of these things are in general  
expected to lead to large changes in the high energy density of  
states.  
  
\noindent  
So far we have discussed the description of the supersymmetric system of  
near-coincident parallel fivebranes.  As for non-singular  
orbifolds \aaa, one expects to find a rich set of non-supersymmetric  
fivebrane configurations with localized tachyons, whose   
condensation has similar effects to those described in previous  
sections. In particular, we argued in section 4 that it should  
be possible to flow from regular orbifolds to geometries with  
throats.   
  
We will not attempt a comprehensive discussion of non-supersymmetric  
throat geometries here, but will give one class of constructions  
where tachyon condensation seems to connect non-supersymmetric CHS type  
geometries to supersymmetric ones.   
  
Consider the background   
\eqn\fivenonsusy{\IR^{5,1}\times   
\left({SL(2)_n\over U(1)}\times {SU(2)_n\over U(1)}\right)/\IZ_{n'}  
}   
where we assume that the integers $n,n'$ are related via:  
\eqn\nnprime{n=n'(2l+1);\;\;l\in \IZ}  
The unperturbed superpotential describing this system is  
\eqn\wunpert{W=\chi^{n'(2l+1)}+\mu e^{-{1\over Q}\Phi}}  
The new phenomenon that occurs here compared to \supgen\ is that  
there exist chiral relevant operators that survive the (chiral)  
GSO projection. They correspond to deformations of the superpotential  
of the form  
\eqn\delW{\delta W=\sum_{j=0}^{l-1}\lambda_j\chi^{n'(2j+1)}}  
The analysis of the perturbed system is elementary in this case,  
since it only involves a relevant deformation of the $N=2$ minimal  
model. One finds a cascade of the form discussed in sections 3,4,   
with a stable endpoint corresponding to $\delta W=\lambda_0\chi^{n'}$.  
As $\lambda_0\to\infty$ in the IR, the system approaches that  
describing $n'$ NS5-branes arranged on a circle; this configuration  
is stable and does not decay further.   
  
This provides an example of a flow from  
a non-supersymmetric to a supersymmetric  
fivebrane system. It is presumably easy to construct many other such   
flows.

 
\newsec{Discussion} 
 
\lref\DabholkarIF{ 
A.~Dabholkar, 
``Tachyon Condensation and Black Hole Entropy,'' 
hep-th/0111004. 
} 
 
The two main results we have found are the following. First, we have 
defined a quantity $g_{\rm cl}$ in theories containing localized closed 
string tachyons which appears to decrease along RG flows that leave 
$c_{\rm eff}$ unchanged. Second, we have shown that $N=2$ world-sheet 
supersymmetry and its associated chiral ring  
is a powerful tool for studying tachyon condensation 
in one class of these theories, namely non-supersymmetric orbifolds 
with localized tachyons. In particular, as in supersymmetric orbifolds, 
the world-sheet CFT captures all the geometrical information of the 
resolved geometry. This is perhaps not surprising since the CFT is 
non-singular and  thus provides in some sense a smoothed out version 
of the geometry, but it is striking how direct the correspondence is. 
 
There are a number of points which should be better understood in 
what we have done. We conjecture that $g_{\rm cl}$ decreases along 
RG flow and have checked this in many examples, but we have no 
proof of this result. A proof of the analogous result in open 
string theory required an off-shell definition 
of $g_{\rm op}$ and the techniques of BSFT \KutasovQP. It would be 
very interesting to see if similar methods could be used to 
define an off-shell version of $g_{\rm cl}$ and a corresponding 
string field theory of the localized closed string states. 
The possibility that $g_{\rm cl}$ should be related to an
action for localized closed string states also has clear 
implications for recent work on closed string tachyons and 
black hole entropy \DabholkarIF. 
 
We also noted earlier that the formula for $g_{\rm cl}$ in orbifold 
CFT is mathematically the same as the formula for the $\eta$ invariant 
for a quotient of $\IC^d$. If this connection can be made more 
precise it would help to explain some of the properties of $g_{\rm cl}$. 
For example it would provide an alternative explanation
of the fact that $g_{\rm cl}$ does not change 
under localized perturbations that do not change the geometry 
at infinity. 
The combinatorics of the Hirzebruch-Jung resolution also encode the 
surgery data for constructing the link of the 
$\IC^2/\IZ_{n(p)}$ singularity. We expect that there 
are several interesting relations of $g_{\rm cl}$ to 
3-manifold invariants and Seiberg-Witten theory. 
 
The relation between geometry and the chiral ring also needs 
to be understood in more detail. In supersymmetric orbifolds, 
the chiral ring can be identified with the (quantum) 
cohomology ring of the resolved geometry.  
In the non-supersymmetric theories  
we are considering, the chiral ring clearly cannot be identified  
directly with the ordinary cohomology of the resolution, because 
the grading given by the $U(1)_R$ charge is not integer, and also 
because some of the relations are incompatible with ordinary 
cohomology.  
 
The chiral ring clearly defines some interesting 
generalization of cohomology.  
In the $\IC^2/\IZ_{n(p)}$ orbifold 
the number of generators in the cohomology defined by the 
chiral ring is the length of the continued fraction $n/p$ 
(or $n/(n+p)$ for $p<0$) defined in section 4. The 
appropriate $K$-theory lattice is expected to be of rank $n$. 
It would be interesting to learn if there is some 
generalization of the McKay correspondence. 
 
A related  issue, raised in  
\refs{\MartinecCF,\AdamsSV}, 
involves the fate of the fractional branes that 
carry these K-theory charges,  
and twisted sector RR gauge fields that measure those charges, 
after the process of tachyon condensation.  
Similar issues arise in studies 
of tachyon condensation in open string theory;  
it may be that some of these questions can be addressed  
more fruitfully in the present context. 
 
Indeed, an interesting issue is the 
interplay of open and closed string dynamics in the presence 
of a closed string tachyon.  A concrete step in this direction 
is the calculation (in appendix B) of the leading corrections 
to the open string dynamics due to an expectation value of a 
twisted sector closed string tachyon.  One may imagine 
that at finite string coupling, the presence of D-branes 
may significantly modify the course of tachyon condensation; 
since the tachyon condensate has the effect of blowing 
up a cycle, and D-brane tension tries to shrink a cycle, 
there should be interesting mechanisms for stabilizing  
cycles due to the competition between these two effects. 
 
\lref\RussoTF{ 
J.~G.~Russo and A.~A.~Tseytlin, 
``Magnetic backgrounds and tachyonic instabilities in closed superstring  theory and M-theory,'' 
Nucl.\ Phys.\ B {\bf 611}, 93 (2001) 
hep-th/0104238. 
} 
 
\lref\GutperleMB{ 
M.~Gutperle and A.~Strominger, 
``Fluxbranes in string theory,'' 
JHEP {\bf 0106}, 035 (2001) 
hep-th/0104136. 
} 
 
\lref\GibbonsPS{ 
G.~W.~Gibbons and K.~I.~Maeda, 
``Black Holes And Membranes In Higher Dimensional Theories With Dilaton Fields,'' 
Nucl.\ Phys.\ B {\bf 298}, 741 (1988). 
} 
 
\lref\DowkerUP{ 
F.~Dowker, J.~P.~Gauntlett, S.~B.~Giddings and G.~T.~Horowitz, 
``On Pair Creation Of Extremal Black Holes And Kaluza-Klein Monopoles,'' 
Phys.\ Rev.\ D {\bf 50}, 2662 (1994) 
hep-th/9312172. 
} 
 
This work also suggests a number of other directions for future research. 
It would  be interesting to generalize our results to 
higher dimensional orbifolds and to orbifolds of the heterotic string. 
It would also be interesting to study 
other systems with localized closed string tachyons.  These include 
orbifolds of $AdS_3$ \MartinecCF, and ``fluxbrane''  solutions in string  
theory \refs{\GibbonsPS,\DowkerUP,\GutperleMB,\RussoTF,\TseytlinQB,
\CostaIF,\SaffinKY,
\SuyamaGD,\RussoNA,\TakayanagiJJ}.


\bigskip  
\noindent{\bf Acknowledgements:}  
We would like to thank  
A. Adams, J. Distler, M. Douglas, E. Diaconescu, D. Friedan, 
J. Maldacena, D. Morrison, J. Polchinski, N. Seiberg, E. Silverstein,   
E. Witten and A. B. Zamolodchikov for useful discussions.    
The support and hospitality of the Aspen Center for Physics  
during the initial stages of this work is gratefully appreciated.
D.K. thanks the Rutgers NHETC for hospitality during part of  
this work. In addition, the work of D.K. and E.M. is  
supported in part by DOE grant DE-FG02-90ER40560; J.H. is  
supported in part by NSF grant PHY-9901194; and the work  
of G.M. is supported in part by DOE grant DE-FG02-96ER40949.

\appendix{A}{Evaluation of $g_{\rm cl}$ for some orbifolds}

\def\vt#1#2#3{ {\vartheta[{#1 \atop  #2}](#3\vert \tau)} }  
  
In this appendix we evaluate certain sums that are needed  
for computing $g_{\rm cl}$ for some of the orbifold CFT's  
discussed in the text. In particular, we focus on the   
case of $\IC^m/\IZ_n$ orbifolds where $\IZ_n$ rotates the   
superfields by $g\cdot Z^i \sim \omega^{p_i}  Z^i$ (with   
$\omega=e^{2 \pi i/n}$) and   
we take the diagonal modular invariant (and spin structure   
sum). The partition sum of the twisted sector states, with   
states weighted by one is  
\eqn\integaa{\ZZ_{\rm tw}(\tau,\bar\tau)=  
\sum_{s=1}^{n-1}{1\over 2n} \sum_{t=0}^{n-1}\sum_{\epsilon_1, \epsilon_2}   
\prod_{i=1}^m  
\biggl\vert {\vt{\epsilon_1 + sp_i/n}{\epsilon_2+tp_i/n}{0}  
\over   
\vt{\half -  sp_i/n}{\half +tp_i/n}{0} }\biggr\vert^2   
\ .}  
Making a modular transformation to $\tau'= -1/\tau$,  the leading terms in the   
$q'$ expansion comes from the terms with $t=\epsilon_2=0$. These give  
\eqn\asympt{  
g(p_1,\dots, p_m;n) =   
{1\over n}  
\sum_{s=1}^{n-1} \prod_{i=1}^m {1\over (2 \sin\pi p_i s/n)^2}   
\ .}  
Curiously, these expressions turn out to be rational   
functions of  $n$. In fact, they can be evaluated explicitly  
by considering contour integrals of the function  
\eqn\contone{ 
f(z)= \cot(\pi n z) 
\prod_{i=1}^m { 1 \over\sin^2 (\pi p_i z) } } 
We will illustrate the method in some special cases.  
 
In all cases we consider contour integrals around  
the contour given by $C_1 - C_2 - C_3 + C_4$ where  
$C_1$ runs along $1-\epsilon + iy $, $-\Lambda\leq y \leq \Lambda$, 
$C_2$ runs along $x + i \Lambda$, $-\epsilon\leq x \leq 1-\epsilon$, 
$C_3$ runs along $-\epsilon + iy $, $-\Lambda\leq y \leq \Lambda$, 
and  
$C_4$ runs along $x - i \Lambda$, $-\epsilon\leq x \leq 1-\epsilon$.  
Here $\epsilon<1/n$.

Let us first consider the case $p_1=\cdots = p_m=1$.   
Simple contour integration gives:  
\eqn\simpone{  
g_m(n):= g(1,1,\dots, 1; n) =   
{1\over n}  
\sum_{s=1}^{n-1}  {1\over (2 \sin\pi s/n)^{2m} }   
= - {\rm Res}_{x=0} {\cot nx\over (2\sin x)^{2m}}   
\ .}  
With this explicit formula we can prove that $g_m(n)$ is   
a strictly increasing function of $n$ for $n\geq 1$ and fixed $m$,  
and thus verify the $g_{cl}$-conjecture for these special  
$C^m/Z_n$ orbifolds as follows. From \simpone\ it follows that   
$g_m(n)$ is of the form $P_m(n)/n$ for some polynomial   
$P_m$ of order $2m$. The polynomial $P_m(n)$ has a root for $n=1$.
Physically, this is because $\IC^m/\IZ_n$ has no localized states
for $n=1$. Mathematically, it follows since
$-2m \cot x (\sin x)^{-2m} = {d\over dx}(\sin x)^{-2m}$.   
We now use the expansions  
\eqn\expdcot{  
\cot x = {1\over x} \biggl(1- \sum_{j=1}^\infty {2^{2j}\vert   
B_{2j}\vert \over (2j)!} x^{2j}\biggr)  
}  
\eqn\expdcoti{ 
{1\over \sin^2 x} = {1\over x^2} \biggl(1+ \sum_{j=1}^\infty {2^{2j}\vert B_{2j}\vert \over (2j)!} (2j-1) x^{2j}\biggr) 
} 
and subtract the expression in \simpone\ from its value at $n=1$ to   
write:   
\eqn\formpm{  
2^{2m} P_m(n) = {\rm Res}_{x=0} \Biggl\{ {1\over x^{2m+1}} \biggl(  
\sum_{j=1}^\infty { \vert B_{2j}\vert \over (2j)!} (n^{2j}-1)(2x)^{2j}\biggr)  
\biggl(1+  
\sum_{j=1}^\infty { (2j-1)\vert B_{2j}\vert \over (2j)!} (2x)^{2j}\biggr)^m  
\Biggr\}  
\ .}  
In particular we have  
\eqn\gees{  
\eqalign{  
g_1(n) & =  {1\over 12} (n-{1\over n} )\cr  
g_2(n) & =  { (n^2+11) (n^2-1)\over 45\cdot 16 n} \ .}  
}  
In general we can  
note that all the coefficients in the series expansions   
in $x$ which appear in \formpm\ are {\it positive}.   
Thus  
\eqn\formpmii{  
g_m(n) = (n-{1\over n})(A_{2m-2} n^{2m-2} + \cdots + A_0)   
\ ,}  
where $A_i$ are positive rational numbers. It follows that $g_m(n)$ is   
strictly increasing for $n\geq 1$.   
  
In the case when the $p_i$ are not all equal we must work harder.  
We will illustrate the method for the case $m=2$ studied in this paper.

For the case $p=\pm 3$ used in the text the method used 
below gives
\eqn\gtwodee{  
\eqalign{  
g(1,3;n)& = {(n^4 + 210 n^2 - 80 n - 291)\over 405 \cdot 16 n}   
	\qquad    n = 2~ mod~ 3\cr  
 & =  {(n^4 + 210 n^2 + 80 n - 291)\over 405  \cdot 16 n}  
	\qquad   n = 1~ mod~ 3\ .}  
}  

The answer for $g(1,p;n)$ for  general $n,p$ 
 is expressed in terms of the continued fraction 
expansion $n/p = [a_1, \dots, a_r]$. Define 
\eqn\defques{
{q_i\over q_{i+1} } = [a_i, a_{i+1},\dots, a_r] = a_i - {q_{i+2}\over q_{i+1} } 
}
where $n=q_1>q_2> q_3 > \cdots > q_{r} = a_r> q_{r+1}=1 $. 

Then the general formula is
\eqn\genform{
g(1,p;n) = n\Biggl[ \sum_{s=1}^r H({q_s\over q_{s+1}}) + \sum_{s=1}^{r-1}  K({q_s\over q_{s+1}})\Biggr]  
}
Here $H\bigl({n\over p} \bigr)$ is defined for fractions in lowest terms with 
\eqn\functone{
H\bigl({n\over p} \bigr) = {1\over 720 n^2 p^2}\biggl( n^4 + 5n^2 p^2 - 3p^4 + 5n^2-5 p^2 - 3\biggr)
}
while $K({n\over p}) $ is defined for fractions in lowest terms with $p>1$ as 
\eqn\functwo{
\eqalign{
K({n\over p}) & := {1\over np} \Biggl[ f(q_2,q_3)+ f(q_3,q_4) 
+ \cdots + f(q_{r-1},q_r)+ f(q_r,1)\Biggr] \cr
f(j,k) & := -{1 \over 360 jk} (j^4 -5j^2 k^2 + k^4 + 3) \cr} 
}

Let us briefly indicate the proof of this result. Evaluating residues 
we get: 
\eqn\sumres{
0 = {\rm Res}_{z=0} \biggl({\cot(\pi n z) \over \sin^2 (\pi z) \sin^2(\pi p z)} \biggr)
+ {1\over \pi } g(1,p;n) + \sum_{t=1}^{p-1} 
 {\rm Res}_{z=t/p} \biggl({\cot(\pi n z) \over \sin^2 (\pi z) \sin^2(\pi p z)} \biggr)
}
{}From this one derives the recursion relation 
\eqn\sumresp{
{1\over q_1} g(1,q_2;q_1)  = {1\over q_2} g(1,q_3;q_2)  
+ H({q_1\over q_2}) + {1\over q_1 q_2} \kappa\bigl({q_1\over q_2}\bigr)
}
where 
\eqn\auxilsum{
\kappa\bigl({n\over p} \bigr)  := {1\over 8 p}  \sum_{t=1}^{p-1} 
    {\cot(\pi t/p ) \cot(\pi nt/p) \over \sin^2 (\pi t/p )  } 
}
This function can, in turn, be evaluated using residues and 
recursion relations. By evaluating residues of the function 
$$\cot \pi z \cot (\pi n z) \cot(\pi p z)/\sin^2(\pi z)$$
along the same contour as above we produce a nice reciprocity formula
\eqn\reciproc{
 \kappa({n\over p} ) + \kappa({p\over n} )+  f(n,p)=0 ,}
where
\eqn\functionthree{
f(n,p) = -{1\over 360}\biggl( n^4 - 5 n^2 p^2 + p^4 + 3\biggr).
}
Equation \reciproc\ is only valid for  $p>1, n>1$ and 
$n,p$ relatively prime. 
Now  note 
that $\kappa({n\over p} )= \kappa({q_1\over q_2} )= - \kappa({q_3\over q_2} )$.
Using this we can evaluate 
\eqn\evalkappa{
\eqalign{
\kappa\bigl({n\over p}\bigr)& = - \kappa\bigl({q_3\over q_2} \bigr)\cr
& = \kappa\bigl({q_2\over q_3} \bigr) + f(q_2,q_3) \cr
& = - \kappa\bigl({q_4\over q_3} \bigr) + f(q_2,q_3) \cr
& =  \kappa\bigl({q_3\over q_4} \bigr) + f(q_3,q_4) + f(q_2,q_3) \cr
& = \cdots \cr}
}
The process terminates when we get to 
\eqn\terminate{
\kappa\bigl({q_{r-1}\over q_r} \bigr)= -\kappa\bigl({1\over q_r} \bigr)= f(q_r,1)
}
This completes the proof.

 
\appendix{B}{Tachyon contributions to  
scalar potentials and gauge couplings}

\def\im{{\rm Im}}  
\def\re{{\rm Re}} 
 
\subsec{Results}

In this appendix we generalize the computation of \DouglasSW\  
of the couplings of twistfields to the scalars and  
gauge fields in a D-brane probe.

The results are as follows: In the $\IC/\IZ_n$ orbifold, with  
$n$ odd, the v.e.v.'s of the twist-fields $\lambda_j$ of the  
$j^{th}$ twisted sector couple to the scalars in the  
D-brane probe Lagrangian as  
\eqn\twistcouple{ 
-8 \pi \sum_{k=1}^n \zeta_k D_k 
} 
where  
\eqn\defzeta{ 
\zeta_k = 
\im\biggl( \sum_{j~ odd} \lambda_j e^{- 2\pi i {jk\over n} } \biggr) 
} 
and  
\eqn\deekay{ 
D_k = \vert Z_{k+1,k}\vert^2 - \vert Z_{k,k-1} \vert^2 
} 
while the coupling to the gauge fields is determined from  
\eqn\gaugecouple{ 
{1\over 4 g_0^2} \sum_{k=1}^n (1+ B_k) F_{\mu\nu}^{(k)} F^{(k), \mu\nu}  
} 
with  
\eqn\beekay{ 
B_k = 4 \pi \re\biggl(\sum_{j=1}^{n-1} \lambda_j e^{- 2\pi i {jk\over n} } \biggr) 
} 

Let us now consider higher-dimensional orbifolds.  
The computation can be repeated for each superfield,  
and need only be done for chiral or antichiral operators, 
for which the fermionic twistfield 
is of the form $e^{i \phi (H-\bar H)}$  
or $e^{i (\phi -1)(H-\bar H)}$, respectively.  
Here $\phi$ is the twist of the superfield under the  
orbifold action. In particular for the   
$\IC^2/\IZ_{n(p)}$ orbifold the results for  
the gauge couplings, \gaugecouple,\beekay\ remain unchanged.  
 
For the purpose of this appendix, it is useful to adopt 
a somewhat different convention for the perturbations 
than is used in the main text.   
The orbifolds $\IC^2/\IZ_{n(p)}$ and $\IC^2/\IZ_{n(-p)}$ 
are isomorphic as CFT's, with (as discussed at the beginning 
of section 4) the isomorphism being the change of complex 
structure $Z_2\to Z_2^*$. 
One could then consider (say) only $p$ positive, but 
allow perturbations that are chiral with respect 
to either one or the other of the two R-symmetries, \uoner\ or 
\eqn\rother{ 
  \tilde J=\psi_1\psi_1^*-\psi_2\psi_2^*\ . 
} 
We denote fields chiral with respect to \uoner\ as  
being of type $(c_1,c_2)$, and fields chiral with respect 
to \rother\ as being of type $(c_1,a_2)$. 
 
The scalar couplings depend on whether the tachyon v.e.v. is  
of $(c_1,c_2)$ type (or its $(a_1,a_2)$ complex conjugate) or of  
$(c_1,a_2)$ type (or its $(a_1,c_2)$ complex conjugate).  
The coupling to $(c_1,c_2)$-type tachyons is of the  
form \twistcouple\ where  
\eqn\zetacc{ 
\zeta_k = 
\im\biggl( \sum_{(c_1,c_2)-type } T_\phi e^{- 2\pi i k\phi } \biggr) 
} 
and  
\eqn\deekay{ 
D_k = \vert Z_{k+1,k}^{(1)}\vert^2 - \vert Z_{k,k-1}^{(1)} \vert^2 
+ \vert Z_{k+p,k}^{(2)}\vert^2 - \vert Z_{k,k-p}^{(2)} \vert^2 
} 
In equation \zetacc\ $\phi$ is the twist of the field $Z^{(1)}$.  
 
Similarly, the coupling to $(c_1,a_2)$-type tachyons  
is of the  
form \twistcouple\ where  
\eqn\zetacc{ 
\zeta_k = 
\im\biggl( \sum_{(c_1,a_2)-type } T_\phi e^{- 2\pi i k\phi } \biggr) 
} 
and  
\eqn\deekay{ 
D_k = \vert Z_{k+1,k}^{(1)}\vert^2 - \vert Z_{k,k-1}^{(1)} \vert^2 
- \vert Z_{k+p,k}^{(2)}\vert^2 +  \vert Z_{k,k-p}^{(2)} \vert^2 
} 

\subsec{Computation}

We need to compute disk amplitudes with one twistfield at the center and  
two vertex operators on the boundary. It suffices to consider the case of  
one twisted $N=2$ scalar superfield $Z$ in a twisted sector  
with $g\cdot Z = e^{2\pi i \phi} Z$.  Here $\phi = \{ {p j\over n} \}$   
for the $j^{th}$ twisted sector of a $\IZ_n$ orbifold when   
the generator of the orbifold group acts as $Z \to \omega^p Z$.

For the coupling to twisted scalars we require the vertex operators 
at momentum $k$:   
\eqn\totvo{ 
\eqalign{ 
\bar Z(k) \otimes V(x;k) & \cr 
 Z(k) \otimes \bar V(x;k) & \cr} 
} 
Here $Z(k)$ is a  matrix valued spacetime field for the  
spacetime coordinate  
$Z$. The field $\bar Z$ is  
the Hermitian conjugate. These fields are   
 valued in the Chan-Paton representation of the  
orbifold group.   $x$ is a coordinate along  
the boundary of the upper half-plane or  
the unit disk.  
The orbifold projection is given by   
$\gamma Z \gamma^{-1} = \omega^p Z$.  
Thus \totvo\ are invariant operators.   
In this paper we are only considering the  
regular representation of the orbifold group,  
appropriate to a D-brane probe,  
so the nonzero entries of $Z(k)$  
are $Z_{r+p,r}$, $1\leq r \leq n$.

Inclusion of the twistfield follows an  
important rule of boundary conformal field 
theory, first  stated in \DouglasSW.  
In boundary conformal field theory  
the twistfield vertex operator $V_\phi(z,\bar z)$ is not mutually  
local with respect to the scalar vertex operators $V(x;k) $ 
and $\bar V(x;k)$.  
Thus, amplitudes with a twistfield in the disk together with   
scalar vertex operators on the boundary 
 are not single valued as functions of $x$. This difficulty  
may be corrected by accompanying the insertion  
of the twistfield with   an insertion of  
the Chan-Paton group action in the combination:  
\eqn\twistfieldop{ 
\gamma_\phi^{-1}\otimes V_\phi 
} 
where $\gamma_\phi Z \gamma_\phi^{-1} = e^{2\pi i \phi} Z$.  
The order of Chan-Paton matrices reflects the order of operators  
relative to the intersection of the cut with the boundary of  
the disk.  
With this insertion, the correlators of boundary vertex operators 
(in a given order)  
\totvo\ are single-valued in the presence of    
the twistfield. In particular, the ordering of the  
boundary operators relative to the cut is irrelevant.  
 
Similarly, the gauge fields have vertex operators  
$A_\mu(k) \otimes V^{\mu}(x;k)$  
where $\gamma_\phi A_\mu(k) \gamma_\phi^{-1} = A_\mu(k)$ 
in the Chan-Paton representation.  
 
The  amplitude connecting the twistfield 
 to the scalars is determined by computing  
the amplitude  
\eqn\dmimprv{ 
\eqalign{ 
T_\phi(k_1)\biggl[{\Tr} \gamma_\phi^{-1} \bar Z(k_2) Z(k_3)   
\int dx_2 dx_3 & 
\langle V_\phi (z,\bar z;\half k_1) V(x_2;k_2)  
	\bar V(x_3;k_3) \rangle + \cr 
+ {\Tr}\gamma_\phi^{-1} Z(k_3) \bar Z(k_2)  \int dx_2 dx_3 & 
\langle V_\phi(z, \bar z;\half k_1) 
	\bar V(x_2;k_3) V(x_3;k_2)  \rangle \Biggr]  \cr} 
} 
Here we are using the upper-half-plane model of the disk.  
We partially fix the Mobius invariance by putting the twistfield at  
$z=i$ and then we integrate 
over $-\infty< x_2< x_3< \infty$. The momenta satisfy  
$k_1+k_2 + k_3=0$.  
 
Similarly, the coupling to the gauge fields is 
\eqn\dmimprvi{ 
\eqalign{ 
T_\phi(k_1)\biggl[{\Tr} \gamma_\phi^{-1} A_\mu(k_2) A_\nu(k_3)   
\int dx_2 dx_3 & 
\langle V_\phi (z,\bar z;\half k_1) V^{\mu}(x_2;k_2)   
	V^{\nu}(x_3;k_3)  \rangle+ \cr 
+ {\Tr}\gamma_\phi^{-1} A_\nu(k_3) A_\mu(k_2) \int dx_2 dx_3 & 
\langle V_\phi(z, \bar z;\half k_1) 
	V^{\nu}(x_2;k_3)  V^{\mu}(x_3;k_2)   \rangle\Biggr]  \cr} 
} 

In order to proceed with the computation we begin by noting that, although  
a priori one must include both orderings  the 
two lines in  \dmimprv\dmimprvi\ make equal contributions (due to  
mutual locality of \totvo\ and \twistfieldop).  
 
We are interested in the coupling of the twist fields $T_\phi(k)$  
at zero momentum. In nonsupersymmetric orbifolds this means  
that this coupling   is an off-shell  
coupling in those sectors where  
$T_\phi$ is massive or tachyonic.  
We will take the fields $Z(k), A_\mu(k)$ to be on-shell, so  
that $k_2^2= k_3^2=0$. Let $\delta:=k_2\cdot k_3 = \half k_1^2$.  
We will compute the vertex operator correlators in the flat  
metric for arbitrary $\delta$ and then   analytically  
continue $\delta \to 0$. This defines our off-shell continuation.  
Technically, it is important to keep the momenta nonvanishing in  
the computation because many terms have canceling zeroes and poles  
in the $\delta \to 0 $ limit.   
 
Let us now describe the vertex operators.  
The gauge field vertex operators will be in the $0$-picture:  
\foot{We suppress all the ghost factors in the text below.}  
\eqn\newgauge{ 
\epsilon_\mu 
V^{\mu}(x;k) = \epsilon_\mu(\p Y^\mu  
	- i   k\cdot \psi \psi^\mu)e^{i k\cdot Y}(x) 
} 
where $Y^\mu$ are the untwisted directions along the brane probe  
and $\psi^\mu$ are the $N=1$ superconformal partners. These satisfy  
Neumann boundary conditions.  
 
The scalar vertex operators are (in the $0$-picture):  
\foot{We use the notation $Z$ for the matrix field,  
the conformal field theory scalar field,  
and the conformal field theory $N=2$ superfield.  
Nevertheless, we hope no confusion will arise.}  
\eqn\newscalar{ 
\eqalign{ 
V(x;k) & = (\p Z -   i k\cdot \psi e^{i H})e^{i k \cdot Y}(x)\cr 
\bar V(x;k) & = (\p \bar Z -   i k\cdot \psi e^{-i H})e^{i k \cdot Y}(x)\cr} 
} 
If the orbifold group acts in the  
$X^8-X^9$ plane then  the complex scalar is  
$Z= {1\over \sqrt{2}}(X^8 + i X^9)$  
and the superconformal partner is bosonized via 
$\psi = {1\over \sqrt{2}}(\psi^8 + i \psi^9) =   e^{i H}$.  
The relative phases in \newscalar, which are crucial, are most readily  
determined by acting with the $N=1$ supercurrent which is proportional to 
\eqn\ennone{ 
G^- + G^+ = \p Z e^{-i H}  + \p \bar Z e^{iH} + \p Y^\mu \psi^\mu  
} 
where  
\eqn\ntwocurr{ 
\eqalign{  
G^- & = \p Z e^{-i H} + \cdots \cr 
G^+  & = \p \bar Z e^{i H} + \cdots \cr 
\bar G^- & = \pb Z e^{-i \bar H}+ \cdots \cr 
\bar G^+  & = \pb \bar Z e^{i\bar H}+ \cdots  \cr} 
} 
The fields $Z$ satisfy Dirichlet boundary conditions  
(hence $B$-type supersymmetry) and hence 
\eqn\dirferm{ 
\bar \psi(\bar z) = - \psi(\bar z) . 
} 
 In terms of the bosonized field $H$:  
\eqn\dirferm{ 
\bar H( \bar z) = H(\bar z) + \pi.  
} 

Finally, the  twistfield vertex operator  
will be taken in the $(-1)$ picture 
\eqn\twistfld{ 
V_\phi(z,\bar z;k) 
  :=   \sigma_{\phi}(z,\bar z) e^{i \phi( H - \bar H)} e^{i k Y}(z,\bar z) 
} 
for $C$-type twistfields and  
\eqn\twistfld{ 
V_\phi(z,\bar z;k) 
  :=   \sigma_{\phi}(z,\bar z) e^{i (\phi-1)( H - \bar H)}  
	e^{i k Y}(z,\bar z) 
} 
for $A$-type twistfields.  
Here $\sigma_\phi$ is  the bosonic twistfield and we take the  
convention of \DixonQV\ that  
\eqn\opetwist{ 
\eqalign{ 
\p Z(z) \sigma_{\phi}(w,\bar w) &\sim   
	(z-w)^{-(1-\phi)} \tau(w,\bar w) \cr} 
} 
where $\tau$ is a single-valued field.  
 
The main twistfield correlator we need is \DixonQV\  
\eqn\threepoint{ 
\eqalign{ 
{\langle \sigma_{\phi}(z,\bar z) \p Z(x_2) \p \bar Z(x_3) \rangle_H  
\over \langle \sigma_{j/n}(z,\bar z) \rangle_H } & = \cr 
=- &  
x_{23}^{-2} ({x_2 - \bar z\over x_2 -  z })^{1-\phi}  
 ({x_3 -  z\over x_3 -  \bar z })^{1-\phi}    
\biggl({\phi} + (1- {\phi}){x_2 -  z\over x_2 -  \bar z } 
{x_3 -  \bar z\over x_3 -   z } \biggr)\cr} 
} 
where $0\leq \phi<1$.  
Recall that we will put $z=i$.  
We choose branch cuts so that  
\eqn\branchcut{ 
w = e^{i \theta} = {x-i \over x+i}  
} 
has argument $0 \leq \theta < 2\pi$. Similarly 
\eqn\scalfer{ 
{\langle  e^{i{\phi}(H-\bar H)} k_2\cdot  
	\psi e^{i H}(x_2) k_3\cdot \psi e^{-i H}(x_3) \rangle 
\over \langle e^{i{j\over n}(H-\bar H)} \rangle }  
= -   {\delta \over x_{23}^2}  
\bigl({x_2-z \over x_2 - \bar z}\bigr)^{\phi}  
\bigl({x_3-\bar z \over  x_3 - z }\bigr)^{\phi } 
} 
We are assuming the cocycles are such that all fermions anticommute,  
so $e^{iH}$ anticommutes with $\psi$.

The amplitude \dmimprvi\ for gauge fields $A^r(k_2) = \epsilon_2$  
and $A^{r'}(k_3) = \epsilon_3$ and all other gauge fields zero  
is  
\eqn\gaugeampl{ 
2\langle V_\phi \rangle e^{i \Phi(\delta)} \delta^{r,r'}  
e^{-2\pi i {r  \phi} } \biggl[  \epsilon_2\cdot \epsilon_3 \delta -  
\epsilon_2\cdot k_3 \epsilon_3\cdot k_2\biggr]  
{\Gamma(\half (\delta+1)) \Gamma(\half) \over \Gamma(\half  
\delta + 1) }  
} 
where $e^{i \Phi(\delta)} \to 1$ as $\delta \to 0$, and  
$\langle V_\phi \rangle$ is the one-point function of the  
twistfield on the disk. We have used the BRST  
conditions on the polarizations. The integral is most easily done by  
fixing $x_2=0$ and integrating $x_3$ along the real axis.

The amplitude \dmimprvi\ for scalar fields is most easily done by  
mapping the amplitude to the integral around the boundary of the  
unit disk using \branchcut. The contour integral can be deformed  
to an integral around a cut in the $w$ plane along $[0,1]$ using:  
\eqn\newint{ 
\eqalign{ 
J(a,b) & := \oint d \xi \xi^a (1-\xi)^b \cr 
& = \int_0^{2\pi} i \xi d \theta e^{i a \theta} (1-e^{i \theta})^b \cr 
& = (e^{2\pi i a} - e^{2\pi i b} )  
	{\Gamma(a+1) \Gamma(b+1)\over \Gamma(a+b+2)} \cr} 
} 
(here $\xi = e^{i \theta} $ in the integrand).  
 
The result is expressed in terms of several gamma functions. Taking  
the limit $\delta \to 0$ results in the amplitude  
\eqn\finalamp{ 
\pm 4 \pi i \langle V_\phi \rangle  
{\Tr} \bigl( \gamma_\phi^{-1} \bar Z Z) (1- e^{-2\pi i \phi})  
} 
where we take the $+$ sign for $C_\phi$ type twistfields and  
the $-$ sign for $A_\phi$-type twistfields. Absorbing  
$\langle V_\phi \rangle $ into the normalization of the  
twistfield $T_\phi(k)$ leads to the D-brane probe  
couplings quoted above.


\appendix{C}{Chiral rings for the marginal deformation 
$\IC^2/\IZ_{12k+2(-3)}\to \IC^2/\IZ_{6k+1(-2)}\oplus\IC^2/\IZ_{6k+1(3k-1)}$} 
 
In this appendix we discuss  the chiral ring of  
\eqn\Cone{\IC^2/\IZ_{2\l(-3)};\;\;\l=6k+1} 
which was discussed in section 4. We also describe the effect on 
the chiral ring of a certain marginal perturbation, which is a 
special case of those described in equation \caseiidaughters. We  
discuss the type 0 version of the theory.  
 
The chiral ring of the model contains the operators 
\eqn\Ctwo{ 
  \sigma^{\sst (1)}_{j\over 2\l} 
	\,e^{i{j\over2\l}(H_1-\bar H_1)} 
 	\,\sigma^{\sst (2)}_{1-\{{3j\over 2\l}\}} 
	\,e^{i(1-\{{3j\over2\l}\})(H_2-\bar H_2)} 
} 
where $j=1,2,\cdots, 2\l-1$; $\{\theta\}$ is the fractional  
part of $\theta$. The R-charges of the operators \Ctwo\ are 
\eqn\Cfour{R_j=\coeff{j}{2\l}+1-\{\coeff{3j}{2\l}\}.} 
Equation \Cfour\ describes three bands: 
\eqn\Cfive{\eqalign{ 
{\rm I}:&\;\; j=1,2,\cdots, 4k\ ;\qquad\qquad R_j=1-\coeff j\l\cr 
{\rm II}:&\;\; j=4k+1,\cdots, 8k+1\ ;\ \quad R_j=2-\coeff j\l\cr 
{\rm III}:&\;\; j=8k+2,\cdots, 12k+1\ ;\quad R_j=3-\coeff j\l\cr 
}} 
The generators of the ring are all operators in the first band, 
$X_j$ with $j=1,2,\cdots, 4k$, as well as the last operator in 
the second band, $X_{8k+1}$. One can check that they satisfy the 
relations \ringrels\ implied by the appropriate continued fraction, 
$[2^{4k-1},3,2]$, \contxb.

The operator with $j=\l=6k+1$ in band II is marginal, and one can 
ask what happens when it is turned on with a large coefficient. 
As shown in section 4, in this limit the CFT is expected to decompose 
into the decoupled factors 
\eqn\Csix{ 
\IC^2/\IZ_{2\l(-3)}\to \IC^2/\IZ_{\l(-2)}\oplus\IC^2/\IZ_{\l(3k-1)},} 
corresponding to the decomposition of the continued fraction 
\eqn\decompcont{[2^{4k-1},3,2]\to [2^{3k-1},3]\oplus [3,2^{k-2},3,2].} 
We next show that the chiral ring of the r.h.s. in \Csix\ agrees 
with \Cfive. Consider first the factor $\IC^2/\IZ_{\l(-2)}$ in \Csix. 
The chiral operators are similar to \Ctwo\ and have R-charges 
\eqn\Cseven{ 
  \tilde R_j={j\over \l}+1-\{\coeff{ 2j}{\l}\}\ ;\qquad j=1,2,\cdots, \l-1. 
} 
They thus arrange into two bands: 
\eqn\Ceight{\eqalign{ 
{\rm \tilde I}:&\;\; j=1,2,\cdots, 3k\ ;\qquad  
	\tilde R_j=1-{\coeff j\l }\cr 
{\rm \widetilde{II}}:&\;\; j=3k+1,\cdots, 6k\ ;\qquad  
	\tilde R_j=2-{\coeff j\l}\cr 
}} 
One can check that the chiral operators \Ceight\ fill out part of the original 
list \Cfive:  
\eqn\Cnine{\eqalign{ 
\tilde R_j&=R_j\quad{\rm for}\ j=1,2,\cdots, 3k\cr 
\tilde R_j&=R_{\l+j}\quad{\rm for}\ j=3k+1,2,\cdots, 6k\cr 
}} 
Thus, the $\IC^2/\IZ_{\l(-2)}$ factor provides the first $(\l-1)/2$ 
operators in band I and the last $(\l-1)/2$ operators in band III 
in \Cfive.  
 
Moving on to the second factor in \Csix, $\IC^2/\IZ_{\l(3k-1)}$, one finds 
chiral operators with R-charges  
\eqn\Cten{ 
   R_j'=\coeff{j}{\l}+\{\coeff{3k-1}{\l}\,j\}\ ;\qquad j=1,2,\cdots, \l-1. 
} 
One can show that \Cten\ precisely completes \Ceight\ to cover all of 
\Cfive\ (except for the marginal operator $R_\l$, which is added to the 
Lagrangian with a large coefficient and decouples): 
\eqn\Celeven{\eqalign{ 
R'_{2j-1}&=R_{3k+j}\cr 
R'_{2j}&=R_{6k+1+j}\cr 
}} 
where $j=1,2,\cdots, 3k$. In verifying \Celeven\ it is useful to note that 
on the first line, for $j=1,2, \cdots, k$, $R_{3k+j}$ belongs to band I 
in \Cfive, while for $j=k+1,\cdots, 3k$ it is in band II. Similarly, on the 
second line of \Celeven, $j=1,2,\cdots, 2k$ gives operators in band II in 
\Cfive, while $j=2k+1, \cdots, 3k$ gives operators in band III.

\listrefs  
\end